\documentclass[preprint,manuscript]{aastex}
\usepackage{apjfonts,natbib}

\def\asca{{\it ASCA\/}}

\def\chandra{{\it Chandra\/}}

\def\hst{{\it {\it HST}\/}}

\def\rosat{{\it ROSAT\/}}

\def\spitzer{{\it Spitzer\/}}

\def\xmm{{\it XMM-Newton\/}}

\def\etal{{et\,al.\,}}

\def\ltsima{$\; \buildrel < \over \sim \;$}
\def\simlt{\lower.5ex\hbox{\ltsima}}
\def\gtsima{$\; \buildrel > \over \sim \;$}
\def\simgt{\lower.5ex\hbox{\gtsima}}
\def\kms{\ifmmode{~{\rm km~s^{-1}}}\else{~km s$^{-1}$}\fi}
\def\lsim{\lower0.3em\hbox{$\,\buildrel <\over\sim\,$}}
\def\gsim{\lower0.3em\hbox{$\,\buildrel >\over\sim\,$}}
\def\hst{{\it HST}}

\def\h2{H$_2$}
\def\flux{erg~cm$^{-2}$~s$^{-1}$}
\def\xlum{erg~s$^{-1}$}

\def\arcsec{\mbox{$^{\prime\prime}$}}
\def\arcmin{\mbox{$^\prime$}}

\def\Lx{$L_{\mbox{\scriptsize{X}}}$}
\def\nsources{776}
\def\nemseven{655}
\def\nemeight{596}
\def\nunique{762}
\def\ntot{915}
\def\ncdfsunique{173}
\def\nnew{589}
\def\nwfi{640}
\def\fbexp{217}
\def\sbexp{216}
\def\hbexp{212}
\def\fbmean{26.3} % x 10^-15 ergs cm^-2 s^-1
\def\sbmean{6.8}
\def\hbmean{34.6}
\def\fblimit{3.5 $\times$ 10$^{-16}$}
\def\sblimit{1.1 $\times$ 10$^{-16}$}
\def\hblimit{6.7 $\times$ 10$^{-16}$}
\def\minsbcnts{4}
\def\minsbflux{8.9~$\times$~10$^{-17}$}

\def\minhbflux{4.4~$\times$~10$^{-16}$}


\shortauthors{LEHMER ET AL.}
\shorttitle{Catalog of the Extended Chandra Deep Field-South}

\begin{document}

%
%%%%%%%%%%%%%%%%%%%%%%%%%%%%%%%%%%%%%%%%%%%%%%%%%%%%%%%%%%%%%%%%%%%%%%%%%%%%%%%%%%
\title{The Extended {\it Chandra} Deep Field-South Survey. {\it Chandra} Point-Source Catalogs}
%%%%%%%%%%%%%%%%%%%%%%%%%%%%%%%%%%%%%%%%%%%%%%%%%%%%%%%%%%%%%%%%%%%%%%%%%%%%%%%%%%
%

\author{
B.~D.~Lehmer,\altaffilmark{1}
W.~N.~Brandt,\altaffilmark{1}
D.~M.~Alexander,\altaffilmark{2}
F.~E.~Bauer,\altaffilmark{3}
D.~P.~Schneider,\altaffilmark{1}
P.~Tozzi,\altaffilmark{4}
J.~Bergeron,\altaffilmark{5}
G.~P.~Garmire,\altaffilmark{1}
R.~Giacconi,\altaffilmark{6}
R.~Gilli,\altaffilmark{7}
G.~Hasinger,\altaffilmark{8}
A.~E.~Hornschemeier,\altaffilmark{9,6}
A.~M.~Koekemoer,\altaffilmark{10}
V.~Mainieri,\altaffilmark{8}
T.~Miyaji,\altaffilmark{11}
M.~Nonino,\altaffilmark{4}
P.~Rosati,\altaffilmark{12}
J.~D.~Silverman,\altaffilmark{8}
G.~Szokoly,\altaffilmark{8}
\& C.~Vignali\altaffilmark{13}
}

\altaffiltext{1}{Department of Astronomy \& Astrophysics, 525 Davey Lab,
The Pennsylvania State University, University Park, PA 16802, USA}
\altaffiltext{2}{Institute of Astronomy, Madingley Road, Cambridge, CB3 0HA, United Kingdom}
\altaffiltext{3}{Columbia Astrophysics Laboratory, Columbia University, Pupin Labortories, 550 W. 120th St., Rm 1418, New York, NY 10027, USA}
\altaffiltext{4}{INAF - Osservatorio Astronomico di Trieste, via G. B. Tiepolo 11, 34131 Trieste, Italy}
\altaffiltext{5}{Institut d'Astrophysique de Paris, 98bis Boulevard, F-75014 Paris, France}
\altaffiltext{6}{Department of Physics and Astronomy, Johns Hopkins University, 3400 North Charles Street, Baltimore, MD 21218, USA}
\altaffiltext{7}{Istituto Nazionale di Astrofisica (INAF) - Osservatorio Astrofisico di Arcetri, Largo E. Fermi 5, 50125 Firenze, Italy}
\altaffiltext{8}{Max-Planck-Institut f\"ur extraterrestrische Physik, Giessenbachstrasse, D-85748 Garching b. M\"unchen, Germany}
\altaffiltext{9}{Laboratory for X-ray Astrophysics, NASA Goddard Space Flight Center, Code 662, Greenbelt, MD 20771, USA}
\altaffiltext{10}{Space Telescope Science Institute, 3700 San Martin Drive, Baltimore, MD 21218, USA}
\altaffiltext{11}{Department of Physics, Carnegie Mellon University, Pittsburgh, PA 15213, USA}
\altaffiltext{12}{European Southern Observatory, Karl-Schwarzschild-Strasse 2, Garching, D-85748, Germany}
\altaffiltext{13}{Dipartimento di Astronomia, Universit\'a degli Studi di Bologna, Via Ranzani 1, 40127 Bologna, Italy}

%
%%%%%%%%%%%%%%%%%%%%%%%%%%%%%%%%%%%%%%%%%%%%%%%%%%%%%%%%%%%%%%%%%%%%%%%%%%%%%%%%%%
\begin{abstract}
%%%%%%%%%%%%%%%%%%%%%%%%%%%%%%%%%%%%%%%%%%%%%%%%%%%%%%%%%%%%%%%%%%%%%%%%%%%%%%%%%%
%

We present \chandra\ point-source catalogs for the Extended {\it Chandra} Deep
Field-South (\hbox{E-CDF-S}) survey.  The \hbox{E-CDF-S} consists of four
contiguous 250~ks {\it Chandra} observations covering an approximately square
region of total solid angle $\approx$0.3 deg$^2$, which flank the existing
$\approx$1~Ms {\it Chandra} Deep Field-South (CDF-S).  The survey reaches
sensitivity limits of $\approx$\sblimit\ \flux\ and $\approx$\hblimit\ \flux\
for the \hbox{0.5--2.0~keV} and \hbox{2--8~keV} bands, respectively.  We detect
\nunique\ distinct \hbox{X-ray} point sources within the \hbox{E-CDF-S}
exposure; \nnew\ of these sources are new (i.e., not previously detected in the
$\approx$1~Ms \hbox{CDF-S}).  This brings the total number of \hbox{X-ray}
point sources detected in the E-CDF-S region to \ntot\ (via the E-CDF-S and
$\approx$1~Ms CDF-S observations).  Source positions are determined using
matched-filter and centroiding techniques; the median positional uncertainty is
$\approx$0\farcs35.  The basic \hbox{X-ray} and optical properties of these
sources indicate a variety of source types, although absorbed active galactic
nuclei (AGNs) seem to dominate.  In addition to our main {\it Chandra} catalog,
we constructed a supplementary source catalog containing 33 lower significance
\hbox{X-ray} point sources that have bright optical counterparts ($R$~$<$~23).
These sources generally have \hbox{X-ray}-to-optical flux ratios expected for
normal and starburst galaxies, which lack a strong AGN component.  We present
basic number-count results for our main \chandra\ catalog and find good
agreement with the $\approx$1~Ms \hbox{CDF-S} for sources with
\hbox{0.5--2.0~keV} and \hbox{2--8~keV} fluxes greater than $3 \times
10^{-16}$~\flux\ and $1 \times 10^{-15}$~\flux, respectively.  Furthermore,
three extended sources are detected in the \hbox{0.5--2.0~keV} band, which are
found to be likely associated with galaxy groups or poor clusters at $z \approx
0.1-0.7$; these have typical rest-frame \hbox{0.5--2.0~keV} luminosities of
(1--5)~$\times$~\hbox{10$^{42}$~\xlum}.

\end{abstract}
%%%%%%%%%%%%%%%%%%%%%%%%%%%%%%%%%%%%%%%%%%%%%%%%%%%%%%%%%%%%%%%%%%%%%%%%%%%%%%%%%%


%
%%%%%%%%%%%%%%%%%%%%%%%%%%%%%%%%%%%%%%%%%%%%%%%%%%%%%%%%%%%%%%%%%%%%%%%%%%%%%%%%%%
\section{Introduction}
%%%%%%%%%%%%%%%%%%%%%%%%%%%%%%%%%%%%%%%%%%%%%%%%%%%%%%%%%%%%%%%%%%%%%%%%%%%%%%%%%%
%

Deep and wide \hbox{X-ray} surveys indicate that the cosmic \hbox{X-ray}
background is largely due to accretion onto supermassive black holes (SMBHs)
integrated over cosmic time (e.g., see Brandt \& Hasinger 2005 for a review).
Follow-up studies of deep-survey sources with \hbox{8--10~m} optical telescopes
as well as multiwavelength correlative studies have shown that most of the
\hbox{X-ray} sources are active galactic nuclei (AGNs), many of which are
obscured (e.g., Bauer et~al. 2004; Szokoly et~al. 2004; Barger et~al. 2005).
\hbox{X-ray} surveys have found the highest density of AGNs on the sky (up to
$\approx$7200~deg$^{-2}$). In addition to AGNs, the deepest \hbox{X-ray}
surveys have also detected respectable numbers of starburst and normal galaxies
out to cosmologically interesting distances (\hbox{$z\approx 1$}; e.g.,
Hornschemeier et~al. 2003; Bauer et~al. 2004; Norman et~al. 2004).

Presently, the two deepest \hbox{X-ray} surveys are the $\approx 2$~Ms
\chandra\ Deep Field-North (\hbox{CDF-N}; Brandt \etal~2001, hereafter B01;
Alexander et~al.  2003, hereafter A03) and the $\approx 1$~Ms \chandra\ Deep
Field-South (\hbox{CDF-S}; Giacconi et~al. 2002, hereafter G02).  These
$\approx$400~arcmin$^2$ surveys have been performed in regions of sky with
extensive multiwavelength coverage.  They have provided \hbox{50--250} times
the sensitivity of surveys by previous \hbox{X-ray} missions, detecting large
numbers of point sources (584 for the \hbox{CDF-N} and 346 for the
\hbox{CDF-S}; G02; A03) and about a dozen extended groups and poor clusters
(Bauer et~al. 2002; G02).

The X-ray surveys performed to date have explored an impressive amount of the
sensitivity versus solid angle ``discovery space'' (see Figure~1 and Brandt \&
Hasinger 2005). However, one limitation of the present surveys is that there is
only a relatively small amount of sky probed to \hbox{0.5--2~keV} flux levels
of \hbox{(2--50)$\times 10^{-17}$~erg~cm$^{-2}$~s$^{-1}$}, a flux regime where
many obscured AGNs are observed (e.g., Bauer \etal~2004).  As a result, our
understanding of the \hbox{X-ray} universe at these faint fluxes suffers from
limited source statistics and field-to-field variance.  To mitigate this
limitation, the Extended \chandra\ Deep Field-South (\hbox{E-CDF-S}) survey was
undertaken as part of the \chandra\ Cycle~5 guest observer program. The
\hbox{E-CDF-S} is composed of four 250~ks \chandra\ ACIS-I pointings flanking
the original \hbox{CDF-S}; these are arranged in a contiguous two-by-two
pattern and cover a total solid angle of $\approx$1100~arcmin$^2$.\footnote{The
$\approx$1~Ms \hbox{CDF-S} data cover $\approx 35$\% of the \hbox{E-CDF-S};
much of this coverage, however, has limited sensitivity due to point spread
function (PSF) broadening and vignetting at large off-axis angles (see \S~3 for
details). The same effects limit the sensitivity and positions derived from the
\xmm\ data (Streblyanska et~al. 2004) extending outside the region with
\chandra\ coverage.} The pointings have sufficient sensitivity to detect the
\hbox{X-ray} emission from moderate-luminosity AGNs (\hbox{$L_{\rm
X}=10^{43}$--$10^{44}$~erg~s$^{-1}$}) to \hbox{$z\approx$~3--6} as well as
\hbox{X-ray} luminous starburst galaxies to $z\approx 1$. The \hbox{E-CDF-S}
therefore can significantly improve understanding of SMBH accretion at high
redshift where the source statistics are still limited.  The contiguous nature
of the E-CDF-S will allow wider field studies of the remarkable AGN clustering
already found in the \hbox{CDF-S} (e.g., Gilli et~al. 2003, 2005), and
comparisons with other surveys of comparable depth (e.g., Stern et~al. 2002;
Harrison et~al. 2003; Wang et~al. 2004a,b; Nandra et~al. 2005) will allow
further assessment of the field-to-field variance of \hbox{X-ray} source
populations.

The \hbox{E-CDF-S} field was selected for this program primarily due to its
superb and growing multiwavelength coverage over a $\approx$900 arcmin$^2$
area, which ensures that it will remain a prime survey field in coming decades
(see Figure~2).  For example, the \hbox{E-CDF-S} has been imaged intensively
with the \hst\ Advanced Camera for Surveys (ACS) via the Galaxy Evolution from
Morphology and Spectral Energy Distributions (GEMS; Rix et~al. 2004; 117 \hst\
orbits) and Great Observatories Origins Deep Survey (GOODS; Giavalisco et~al.
2004; 199 \hst\ orbits) projects.  Excellent ground-based imaging is also
available (e.g., Arnouts et~al. 2001; Renzini et~al. 2003; Giavalisco et~al.
2004; Wolf et~al. 2004; Gawiser et~al. 2005), and several spectroscopic
campaigns are underway to identify sources in the \hbox{E-CDF-S}, most notably
with the Very Large Telescope (VLT; e.g., Le~Fevre et~al. 2004; Szokoly et~al.
2004; Vanzella et~al. 2005).  The \hbox{E-CDF-S} has been targeted by \spitzer\
via the GOODS (M. Dickinson et~al., in preparation), the \spitzer\ Wide-Area
Infrared Extragalactic Survey (SWIRE; Lonsdale et~al. 2003), guaranteed time
(e.g., Papovich et~al. 2004), and guest observer (PI: P.~van~Dokkum) programs.
Radio observations of the \hbox{E-CDF-S} have been made with the Australia
Telescope Compact Array (ATCA; J. Afonso et~al., in preparation) and the Very
Large Array.

In this paper, we present \chandra\ point-source catalogs and data products
derived from the \hbox{E-CDF-S} data set along with details of the
observations, data reduction, and technical analysis. The observational
procedures and data processing were similar in nature to those presented in B01
and A03.  Detailed follow-up investigations and scientific interpretation of
the \hbox{E-CDF-S} sources will be presented in subsequent papers.

The Galactic column density along the line of sight to the \hbox{E-CDF-S} is
remarkably low: \hbox{$N_{\rm H}=8.8\times 10^{19}$~cm$^{-2}$} (e.g., Stark
et~al.  1992).  The coordinates throughout this paper are J2000.  Cosmological
parameters of $H_0=70$~km~s$^{-1}$~Mpc$^{-1}$, $\Omega_{\rm M}=0.3$, and
$\Omega_{\Lambda}=0.7$ are adopted.

%
%%%%%%%%%%%%%%%%%%%%%%%%%%%%%%%%%%%%%%%%%%%%%%%%%%%%%%%%%%%%%%%%%%%%%%%%%%%%%%%%%%
\section{Observations and Data Reduction}
%%%%%%%%%%%%%%%%%%%%%%%%%%%%%%%%%%%%%%%%%%%%%%%%%%%%%%%%%%%%%%%%%%%%%%%%%%%%%%%%%%
%

\subsection{Instrumentation and Observations}

The Advanced CCD Imaging Spectrometer (ACIS; Garmire et~al. 2003) was used for
all of the \chandra\ observations.\footnote{For additional information on ACIS
and \chandra\ see the \chandra\ Proposers' Observatory Guide at
http://cxc.harvard.edu/proposer/CfP/.} ACIS is composed of ten CCDs (each
$1024\times1024$ pixels) designed for efficient \hbox{X-ray} source detection
and spectroscopy. ACIS-I consists of four CCDs (CCDs \hbox{I0--I3}) arranged in
a $2\times 2$ array with each CCD tipped slightly to approximate the curved
focal surface of the \chandra\ High Resolution Mirror Assembly (HRMA). The
aim-point of ACIS-I lies on CCD I3.  The remaining six CCDs (ACIS-S; CCDs
\hbox{S0--S5}) reside in a linear array and are tipped to approximate the
Rowland circle of the objective gratings that can be inserted behind the HRMA.

The ACIS-I full field of view is $16\farcm 9\times 16\farcm 9$ ($\approx$285
arcmin$^2$), and the sky-projected ACIS pixel size is $\approx$$0\farcs 492$.
The PSF is smallest at the lowest photon energies and for sources at small
off-axis angles. For example, the 95\% encircled-energy radius at 1.5~keV for
off-axis angles of \hbox{$0^{\prime}$--$8^{\prime}$} is \hbox{$\approx 1\farcs
8$--$7\farcs5$} (Feigelson, Broos, \& Gaffney 2000; Jerius et~al.
2000).\footnote{Feigelson et~al. (2000) is available at
http://www.astro.psu.edu/xray/acis/memos/memoindex.html.} The PSF is
approximately circular at small off-axis angles, broadens and elongates at
intermediate off-axis angles, and becomes complex at large off-axis angles. 

The entire {\it Chandra} observation program consisted of nine separate
\chandra\ observations taken between 2004 February 29 and 2004 November 20 and
is described in Table~1.  The four \hbox{ACIS-I} CCDs were operated in all of
the observations; the ACIS-S CCD S2 was in operation for observations
\hbox{5019--5022} and 6164.  Due to the large off-axis angle of ACIS-S, and
consequently its low sensitivity, these data were not used in this analysis.
All observations were taken in Very Faint mode to improve the screening of
background events and thus increase the sensitivity of ACIS in detecting faint
\hbox{X-ray} sources.\footnote{For more information on the Very Faint mode see
http://cxc.harvard.edu/cal/Acis/Cal\_prods/vfbkgrnd/ and Vikhlinin (2001).} The
observations were made in four distinct observational fields (hereafter, fields
1, 2, 3, and 4; see Table~1 for more observational details) and cover a total
solid angle of 1128.4 arcmin$^2$.  The focal-plane temperature was kept at
$\approx -120^\circ$\,C for all of the nine observations.  

Background light curves for all nine observations were inspected using {\sc
event browser} in the Tools for ACIS Real-time Analysis ({\sc tara}; Broos
et~al. 2000) software package.\footnote{{\sc tara} is available at
http://www.astro.psu.edu/xray/docs.} All but two are free from significant
flaring and are stable to within $\approx$20\%. The two observations with
significant flaring are 5015 and 5017.  The background was $\simgt$1.5 times
higher than nominal for two $\approx$1~ks intervals of observation 5015, and
during observation 5017 the background rose to $\simgt$1.5 times the nominal
rate and remained above this level for $\approx$10~ks near the end of the
observation.  Intervals with flaring were retained because the flaring
strengths were not strong enough to have significant negative effects on our
analyses.

\subsection{Data Reduction}

\chandra\ \hbox{X-ray} Center (hereafter CXC) pipeline software was used for basic
data processing, and the pipeline versions are listed in Table~1.  The
reduction and analysis of the data used \chandra\ Interactive Analysis of
Observations ({\sc ciao}) Version~3.2 tools whenever possible;\footnote{See
http://cxc.harvard.edu/ciao/ for details on {\sc ciao}.} however, custom
software, including the {\sc tara} package, was also used extensively.

All data were corrected for the radiation damage sustained by the CCDs during
the first few months of \chandra\ operations using the Charge Transfer
Inefficiency (CTI) correction procedure of Townsley et~al. (2000,
2002).\footnote{The software associated with the correction method of Townsley
et~al. (2000, 2002) is available at
http://www.astro.psu.edu/users/townsley/cti/.}  In addition to correcting
partially for the positionally dependent grade distribution due to CTI effects,
this procedure also partially corrects for quantum efficiency losses (see
Townsley et~al. 2000, 2002 for further details).

All bad columns, bad pixels, and cosmic ray afterglows were removed using the
``status'' information in the event files, and we only used data taken during
times within the CXC-generated good-time intervals. The {\sc ciao} tool {\sc
acis\_process\_events} was used to remove the standard pixel randomization.

%
%%%%%%%%%%%%%%%%%%%%%%%%%%%%%%%%%%%%%%%%%%%%%%%%%%%%%%%%%%%%%%%%%%%%%%%%%%%%%%%%%%
\section{Production of the Point-Source Catalogs}
%%%%%%%%%%%%%%%%%%%%%%%%%%%%%%%%%%%%%%%%%%%%%%%%%%%%%%%%%%%%%%%%%%%%%%%%%%%%%%%%%%
%

By design, the four 250~ks \hbox{E-CDF-S} observations have their regions of
highest sensitivity located where the sensitivity of the original $\approx$1~Ms
\hbox{CDF-S} observation is poorest (see Table~1 and Figures~2 and 17). The
loss of sensitivity is due to the combination of substantial degradation of the
\chandra\ PSF at large off-axis angles and vignetting. In fact, most of the
area where the \hbox{E-CDF-S} observations have their highest sensitivity lack
any \chandra\ coverage in the $\approx$1~Ms \hbox{CDF-S}.  We experimented with
source searching utilizing the addition of the $\approx$1~Ms \hbox{CDF-S} and
the 250~ks \hbox{E-CDF-S} images; such searching was done with {\sc wavdetect}
(Freeman et~al. 2002) runs that did not utilize detector-specific PSF
information (i.e., the ``DETNAM'' keyword in the image files was deleted).
However, such searching did not find a substantial number of new sources
compared to those presented below combined with those from the original
$\approx$1~Ms \hbox{CDF-S} (G02; A03); these results were verified via
inspection of adaptively smoothed images.  Therefore, our basic approach here
is to present just the sources detected in the new 250~ks \hbox{E-CDF-S}
observations.  The X-ray sources in the $\approx$1~Ms \hbox{CDF-S} catalog of
A03 were processed using the same techniques presented here with two main
differences:

\begin{enumerate}

\item Our main \chandra\ catalog includes sources detected by running {\sc
wavdetect} at a false-positive probability threshold of 10$^{-6}$, somewhat less
conservative than the 10$^{-7}$ value adopted by A03; see $\S$~3.2 for
details.

\item The \hbox{E-CDF-S} consists of four ACIS-I observational fields and
subtends a larger solid angle than the fields presented in
A03.  Therefore, our main \chandra\ catalog of the entire \hbox{E-CDF-S}
exposure was generated by merging sub-catalogs created in each of the four
observational fields; see $\S$~3.2, Table~1, and Figure~2 for details.

\end{enumerate}

\subsection{Image and Exposure-Map Creation}

We constructed images of each of the four E-CDF-S fields using the standard
\asca\ grade set (\asca\ grades 0, 2, 3, 4, 6) for three standard bands (i.e.,
12 images in total): \hbox{0.5--8.0~keV} (full band; FB), \hbox{0.5--2.0~keV}
(soft band; SB), and \hbox{2--8~keV} (hard band; HB).  These images have
0\farcs492 per pixel.  For each of the standard bands, the images from all four
observational fields were merged into a single image using the {\sc ciao}
script {\sc merge\_all}.\footnote{See
http://cxc.harvard.edu/ciao/threads/merge\_all/}  In Figures~3 and 4 we display
the full-band raw and exposure-corrected adaptively smoothed images (see
discussion below), respectively.\footnote{Raw and adaptively smoothed images
for all three standard bands are available at the \hbox{E-CDF-S} website (see
http://www.astro.psu.edu/users/niel/ecdfs/ecdfs-chandra.html).  Furthermore,
equivalent images obtained by merging the \hbox{E-CDF-S} and \hbox{CDF-S} are
also available at the \hbox{E-CDF-S} website.}  Our point-source detection
analyses have been restricted to the raw images constructed for each of the
four observational fields so that the \chandra\ PSF is accounted for correctly
(see $\S$~3.2).

We constructed exposure maps for each of the four observational fields in the
three standard bands.  These were created following the basic procedure
outlined in $\S$~3.2 of Hornschemeier et al.~(2001) and are normalized to the
effective exposures of sources located at the aim points.  Briefly, this
procedure takes into account the effects of vignetting, gaps between the CCDs,
bad column filtering, and bad pixel filtering.  Also, with the release of {\sc
ciao} version 3.2, the spatially dependent degradation in quantum efficiency
due to contamination on the ACIS optical blocking filters is now incorporated
into the generation of exposure maps.\footnote {See
http://cxc.harvard.edu/ciao/why/acisqedeg.html}  A photon index of
$\Gamma=1.4$, the slope of the \hbox{X-ray} background in the
\hbox{0.5--8.0~keV} band (e.g.,\ Marshall \etal 1980; Gendreau \etal 1995), was
assumed in creating the exposure maps.  For each standard band, a total
exposure map, covering the entire \hbox{E-CDF-S}, was constructed by merging
the exposure maps of the four observational fields using the {\sc ciao} script
{\sc dmregrid}.  The resulting full-band exposure map is shown in Figure~5.
Figure~6 displays the survey solid angle as a function of full-band effective
exposure for both the total \hbox{E-CDF-S} exposure (Figure~6a) and the four
individual observational fields (Figure~6b).  Each observational field has
comparable coverage with the majority of the solid angle coverage
($\approx$900~arcmin$^2$) having at least 200~ks of effective exposure.

Using the exposure maps and adaptively smoothed images discussed above, we
produced exposure-corrected images following the prescription outlined in
$\S$~3.3 of Baganoff et~al. (2003).  Figure~7 shows a ``false-color'' composite
image made using exposure-corrected adaptively smoothed \hbox{0.5--2.0~keV}
(red), \hbox{2--4~keV} (green), and \hbox{4--8~keV} (blue) images.

\subsection{Point-Source Detection}

Point-source detection was performed in each band with {\sc wavdetect} using a
``$\sqrt{2}$~sequence'' of wavelet scales (i.e.,\ 1, $\sqrt{2}$, 2,
$2\sqrt{2}$, 4, $4\sqrt{2}$, and 8 pixels).  Our key criterion for source
detection, and inclusion in the main \chandra\ catalog, is that a source must
be found with a given false-positive probability threshold in at least one of
the three standard bands. The false-positive probability threshold in
each band was set to $1\times 10^{-6}$; a total of \nunique\ distinct sources met this
criterion.  We also ran {\sc wavdetect} using false-positive probability
thresholds of $1\times 10^{-7}$ and $1\times 10^{-8}$ to evaluate the
significance of each detected source.

If we conservatively treat the 12 images (i.e., the three standard bands over
the four observational fields) as being independent, it appears that
$\approx$50 (i.e., $\approx$6\%) false sources are expected in our total {\it
Chandra} source catalog for the case of a uniform background over
$\approx$5.0~$\times$~10$^7$ pixels.  However, since {\sc wavdetect} suppresses
fluctuations on scales smaller than the PSF, a single pixel usually should not
be considered a source detection cell, particularly at large off-axis angles.
Hence, our false-source estimates are conservative.  As quantified in
$\S$~3.4.1 of A03 and by new source-detection simulations (P.~E.~Freeman 2005,
private communication), the number of false-sources is likely
\hbox{$\approx$2--3} times less than our conservative estimate, leaving only
\hbox{$\approx$15--25} (i.e., $\simlt$3\%) false sources.  In $\S$~3.3.1 below
we provide additional source-significance information that a user can utilize
to perform more conservative source screening if desired.

\subsection{Point-Source Catalogs}

\subsubsection{Main \chandra\ Source Catalog}

We ran {\sc wavdetect} with a false-positive probability threshold of $1\times
10^{-6}$ on all of the 12 images. The resulting source lists were then merged
to create the point-source catalog given in Table~2. For cross-band
matching, a matching radius of $2\farcs 5$ was used for sources within
$6\arcmin$ of the average aim point. For larger off-axis angles, a matching
radius of $4\farcs 0$ was used.  These matching radii were chosen based on
inspection of histograms showing the number of matches obtained as a function
of angular separation (e.g., see \S2 of Boller et~al. 1998); with these radii the
mismatch probability is $\simlt$1\% over the entire field.

We improved the {\sc wavdetect} source positions using a matched-filter
technique (A03).  This technique convolves the full-band image in the vicinity
of each source with a combined PSF. The combined PSF is automatically generated
as part of the {\sc acis\_extract} procedure (Broos \etal 2002) within {\sc
tara} (see Footnote~5) and is produced by combining the ``library'' PSF of a
source for each observation, weighted by the number of detected
counts.\footnote{{\sc acis\_extract} can be accessed from
http://www.astro.psu.edu/xray/docs/TARA/ae\_users\_guide.html. The PSFs are
taken from the CXC PSF library; see
http://cxc.harvard.edu/ciao/dictionary/psflib.html.} This technique takes into
account the fact that, due to the complex PSF at large off-axis angles, the
\hbox{X-ray} source position is not always located at the peak of the
\hbox{X-ray} emission.  The matched-filter technique provides a small
improvement ($\approx$0$\farcs$1 on average) in the positional accuracy for
sources further than 6$\arcmin$ from the average aim-point.  For sources with
off-axis angles ($\theta$)~$<$~6\arcmin, we found that the
off-axis-angle-weighted combination of centroid and matched-filter positions
returned the most significant improvement to source positions.  Algebraically,
this can be written as:

\begin{equation}
(6\arcmin-\theta)/6\arcmin \times \mbox{ centroid position } + \; \theta/6\arcmin \times \mbox{ matched-filter position } 
\end{equation}

\noindent  This method is similar to that employed by A03.

Manual correction of the source properties was required in some special cases:
(1) There were 11 close doubles (i.e., sources with overlapping PSFs) and one
close triple.  These sources incur large photometric errors due to the
difficulty of the separation process.  (2) A total of eight sources were located
close to bright sources, in regions of high background, in regions with strong
gradients in exposure time, or partially outside of an observational field.  
The properties of these sources have been adjusted manually and
are flagged in column~39 of Table~2 (see below). 

For each observational field, we refined the absolute \hbox{X-ray} source
positions by matching \hbox{X-ray} sources from the main point-source catalog
to $R$-band optical source positions from deep observations ($R_{{\rm lim, } 6
\sigma}$~$\approx$~27 [AB] over the entire E-CDF-S) obtained with the Wide
Field Imager (WFI) of the MPG/ESO telescope at LaSilla (see $\S$~2 of
Giavalisco \etal~2004).  \hbox{X-ray} sources from each of the four
observational fields were matched to optical sources using a 2$\farcs$5
matching radius.  Using this matching radius, a small number of sources were
observed to have more than one optical match; the brightest of these sources
was selected as the most probable counterpart.  Under these criteria, \nwfi\
($\approx$84\%) \hbox{X-ray} sources have optical counterparts.  We also note
that in a small number of cases the \hbox{X-ray} source may be offset from the
center of the optical source even though both are associated with the same
galaxy (e.g., a galaxy with bright optical emission from starlight that also
has an off-nuclear ultraluminous \hbox{X-ray} binary with
\Lx~$\approx$~10$^{38-40}$~\xlum; see e.g., Hornschemeier et~al. 2004).  The
accuracy of the \hbox{X-ray} source positions was improved by centering the
distribution of offsets in right ascension and declination  between the optical
and X-ray source positions; this resulted in small ($<$~1\farcs0)
field-dependent astrometric shifts for all sources in each field.  We also
checked for systematic offsets as a function of right ascension and declination
that may arise from differing ``plate scales'' and rotations between the X-ray
and optical images.  These investigations were performed by plotting the right
ascension and declination offsets (between optical and X-ray sources) as
functions of right ascension and declination; no obvious systematic offsets
were found.  

Figure~8 shows the positional offset between the \hbox{X-ray} and optical
sources versus off-axis angle after applying the positional corrections
discussed above.  Here, the off-axis angles are computed for each observational
field appropriately; this allows for the consistent analysis of {\it Chandra}
positional uncertainties as a function of off-axis angle.  The median offset is
$\approx$0$\farcs$35; however, there are clear off-axis angle and source-count
dependencies. The off-axis angle dependence is due to the HRMA PSF becoming
broad at large off-axis angles, while the count dependency is due to the
difficulty of centroiding a faint \hbox{X-ray} source.  The median offset of
the bright \hbox{X-ray} sources ($\ge$~50 full-band counts) is only
$\approx$0$\farcs$25, while the median offset of the faint \hbox{X-ray} sources
($<$~50 full-band counts) is $\approx$0$\farcs$47.  The positional uncertainty
of each source is estimated following equations 2 and 3.

The main {\it Chandra} source catalog is presented in Table~2, and the details
of the columns are given below.

\begin{itemize}

\item
Column~1 gives the source number. Sources are listed in order of
increasing RA.

\item Columns~2 and 3 give the RA and Dec of the \hbox{X-ray} source,
respectively.  Note that more accurate positions are available for sources
detected near the aim-point of the $\approx$1~Ms CDF-S through the catalogs
presented in A03; see columns~19--21.  To avoid truncation error, we
quote the positions to higher precision than in the International Astronomical
Union (IAU) registered names beginning with the acronym ``CXO ECDFS'' for
``\chandra\ \hbox{X-ray} Observatory Extended \chandra\ Deep Field-South.''
The IAU names should be truncated after the tenths of seconds in RA and after
the arcseconds in Dec.

\item Column~4 gives the positional uncertainty. As shown above, the positional
uncertainty is dependent on off-axis angle and the number of detected counts.
For the brighter \hbox{X-ray} sources ($\ge$~50 full-band counts) the
positional uncertainties are given by the empirically determined equation:

\begin{equation}
\Delta= \left\{\begin{array}{ll}
0.6 & \theta<5\arcmin \\
 & \\
0.6+\left({\theta-5\arcmin\over 20\arcmin}\right) & \theta\ge5\arcmin \\
\end{array}
\right.
\end{equation}

\noindent
where $\Delta$ is the positional uncertainty in arcseconds and $\theta$ is
the off-axis angle in arcminutes (compare with Figure~8). 

For the fainter \hbox{X-ray} sources ($<$~50 full-band counts) the positional
uncertainties are given by the empirically determined equation:

\begin{equation}
\Delta= \left\{\begin{array}{ll}
0.85 & \theta<5\arcmin \\
 & \\
0.85+\left({\theta-5\arcmin\over 4\arcmin}\right) & \theta\ge5\arcmin \\
\end{array}
\right.
\end{equation}

The stated positional uncertainties are somewhat conservative, corresponding to
the \hbox{$\approx$80--90\%} confidence level.

\item Column~5 gives the off-axis angle for each source in arcminutes. This is
calculated using the source position given in columns~2 and 3 and the aim point
(see Table~1) for the corresponding field in which it was detected (column~37).

\item Columns~6--14 give the source counts and the corresponding $1\sigma$
upper and lower statistical errors (from Gehrels 1986), respectively, for the
three standard bands.  All values are for the standard \asca\ grade set, and
they have not been corrected for vignetting. Source counts and statistical
errors have been calculated using circular aperture photometry; extensive
testing has shown that this method is more reliable than the {\sc wavdetect}
photometry.  The circular aperture was centered at the position given in
columns 2 and 3 for all bands.

The local background is determined in an annulus outside of the
source-extraction region. The mean number of background counts per pixel is
calculated from a Poisson model using ${n_1}\over{n_0}$, where $n_0$ is the
number of pixels with 0 counts and $n_1$ is the number of pixels with 1 count.
Although only the numbers of pixels with 0 and 1 counts are measured, this
technique directly provides the mean background even when ${n_1}\gg{n_0}$.
Furthermore, by ignoring all pixels with more than 1~count, this technique
guards against background contamination from sources.  We note that relatively
bright nearby sources may contribute counts to nearby pixels where the
background is estimated.  Since the number density of relatively bright sources
in the E-CDF-S is low, we estimate that only \hbox{$\approx$10--20} of these
sources are thereby contaminated; the majority of these sources have been
corrected via manual photometry of close doubles (see above).  The principal
requirement for using this Poisson-model technique is that the background is
low and follows a Poisson distribution; in \S4.2 of A03 it has been shown that
the \hbox{ACIS-I} background matches this criterion for exposures as long as
$\approx$2~Ms. The total background for each source is calculated and
subtracted to give the net number of source counts.

For sources with fewer than 1000 full-band counts, we have chosen the aperture
radii based on the encircled-energy function of the \chandra\ PSF as determined
using the CXC's {\sc mkpsf} software (Feigelson et~al. 2000; Jerius et~al.
2000). In the soft band, where the background is lowest, the aperture radius
was set to the 95\% encircled-energy radius of the PSF. In the other bands, the
90\% encircled-energy radius of the PSF was used. Appropriate aperture
corrections were applied to the source counts by dividing the extracted source
counts by the encircled-energy fraction for which the counts were extracted.

For sources with more than 1000 full-band counts, systematic errors in the
aperture corrections often exceed the expected errors from photon statistics
when the apertures described in the previous paragraph are used. Therefore, for
such sources we used larger apertures to minimize the importance of the
aperture corrections; this is appropriate since these bright sources dominate
over the background. We set the aperture radii to be twice those used in the
previous paragraph and inspected these sources to verify that the
measurements were not contaminated by neighboring objects.

We have performed several consistency tests to verify the quality of the
photometry. For example, we have checked that the sum of the counts measured in
the soft and hard bands does not differ from the counts measured in the full
band by an amount larger than that expected from measurement error.  Systematic
errors that arise from differing full-band counts and soft-band plus hard-band
counts are estimated to be $\simlt$4\%.

When a source is not detected in a given band, an upper limit is calculated;
upper limits are indicated as a ``$-$1'' in the error columns.  All upper limits are
determined using the circular apertures described above.  When the number of
counts in the aperture is $\leq 10$, the upper limit is calculated using the
Bayesian method of Kraft, Burrows, \& Nousek (1991) for 99\% confidence. The
uniform prior used by these authors results in fairly conservative upper limits
(see Bickel 1992), and other reasonable choices of priors do not materially
change our scientific results.  For larger numbers of counts in the aperture,
upper limits are calculated at the $3\sigma$ level for Gaussian statistics.

\item Columns~15 and 16 give the RA and Dec of the optical source centroid,
which was obtained by matching our \hbox{X-ray} source positions (columns~2 and
3) to WFI $R$-band positions using a matching radius of 1.5 times the
positional uncertainty quoted in column~4.  For a small number of sources more
than one optical match was found, and for these sources the brightest match was
selected as the most probable counterpart.  Using these criteria, 594
($\approx$78\%) of the sources have optical counterparts.  Note that the
matching criterion used here is more conservative than that used in the
derivation of our positional errors discussed in $\S$~3.3.1.  Sources with no
optical counterparts have RA and Dec values set to \hbox{``00 00 00.00''} and
\hbox{``+00 00 00.0''}.

\item Column~17 gives the measured offset between the optical and \hbox{X-ray}
sources (i.e., $O-X$) in arcseconds.  Sources with no optical counterparts have
a value set to ``0''.

\item Column~18 gives the $R$-band magnitude (AB) of each \hbox{X-ray} source.
Sources with no optical counterparts have a value set to ``0''.

\item Column~19 gives the $\approx$1~Ms CDF-S source number from the main
\chandra\ catalog presented in A03 (see column~1 of Table~3a in A03) for
\hbox{E-CDF-S} sources that were matched to A03 counterparts.  We used a
matching radius of 1.5 times the sum of the positional errors of the
\hbox{E-CDF-S} and A03 source positions.  We note that for each matched source
only one match was observed; \hbox{E-CDF-S} sources with no A03 match have a
value of ``0''.

\item Columns~20 and 21 give the RA and Dec of the corresponding $\approx$1~Ms
CDF-S A03 source indicated in column~19.  Sources with no A03 match have RA and
Dec values set to \hbox{``00 00 00.00''} and \hbox{``+00 00 00.0''}.

\item Column~22 gives the $\approx$1~Ms CDF-S source number from the main
\chandra\ catalog presented in G02 (see ``ID'' column of Table~2 in G02) for
E-CDF-S sources that were matched to G02 counterparts.  When matching our
\hbox{E-CDF-S} source positions with G02 counterparts, we removed noted offsets
to the G02 positions of $-1\farcs2$ in RA and $+0\farcs8$ in Dec (see $\S$~A3
of A03); these positions are corrected in the quoted source positions in
columns~23 and 24.  We used a matching radius of 1.5 times the \hbox{E-CDF-S}
positional error plus the G02-quoted positional error for each source position.
We note that for each matched source only one match was observed;
\hbox{E-CDF-S} sources with no G02 match have a value of ``0''.

\item Columns~23 and 24 give the RA and Dec of the corresponding $\approx$1~Ms
CDF-S G02 source indicated in column~22.  Note that the quoted positions have
been corrected by the noted offsets described in column~22 (see $\S$~A3 of
A03). Sources with no G02 match have RA and Dec values set to \hbox{``00 00
00.00''} and \hbox{``+00 00 00.0''}.

\item Columns~25--27 give the effective exposure times derived from the
standard-band exposure maps (see \S3.1 for details on the exposure maps).
Dividing the counts listed in columns~6--14 by the corresponding effective
exposures will provide vignetting-corrected and quantum efficiency
degradation-corrected count rates.

\item Columns~28--30 give the band ratio, defined as the ratio of counts
between the hard and soft bands, and the corresponding upper and lower errors,
respectively. Quoted band ratios have been corrected for differential
vignetting between the hard band and soft band using the appropriate exposure
maps. Errors for this quantity are calculated following the ``numerical
method'' described in \S1.7.3 of Lyons (1991); this avoids the failure of the
standard approximate variance formula when the number of counts is small (see
\S2.4.5 of Eadie et~al. 1971).  Note that the error distribution is not
Gaussian when the number of counts is small.  Upper limits are calculated for
sources detected in the soft band but not the hard band and lower limits are
calculated for sources detected in the hard band but not the soft band.  For
these sources, the upper and lower errors are set to the computed band ratio.

\item Columns~31--33 give the effective photon index ($\Gamma$) with upper and
lower errors, respectively, for a power-law model with the Galactic column
density. The effective photon index has been calculated based on the band ratio
in column~28 when the number of counts is not low.

A source with a low number of counts is defined as being (1) detected in the
soft band with $<30$ counts and not detected in the hard band, (2) detected in
the hard band with $<15$ counts and not detected in the soft band, (3) detected
in both the soft and hard bands, but with $<15$ counts in each, or (4) detected
only in the full band.  When the number of counts is low, the photon index is
poorly constrained and is set to $\Gamma=1.4$, a representative value for faint
sources that should give reasonable fluxes.  Upper and lower limits are
indicated by setting the upper and lower errors to the computed effective
photon index.

\item Columns~34--36 give observed-frame fluxes in the three standard bands;
quoted fluxes are in units of $10^{-15}$~erg~cm$^{-2}$~s$^{-1}$.
Fluxes have been computed using the counts in \hbox{columns~6, 9, and 12}, the
appropriate exposure maps (columns~25--27), and the spectral slopes given in column~31. The
fluxes have not been corrected for absorption by the Galaxy or material
intrinsic to the source. For a power-law model with $\Gamma=1.4$, the soft-band
and hard-band Galactic absorption corrections are $\approx$2.1\% and $\approx
0.1$\%, respectively.  More accurate fluxes for these sources would require
direct fitting of the \hbox{X-ray} spectra for each observation, which is
beyond the scope of this paper.

\item Column~37 gives the observational field number corresponding to the
detected source.  The observational fields overlap in a few areas (see
Figures~5 and 6a) over $\approx$50~arcmin$^2$, which allowed for duplicate
detections of a single source.  Fourteen sources in the {\it Chandra}
catalog were detected in more than one observational field; these sources are
flagged in column 39 (see below).  The data from the observation that produced the greatest
number of full-band counts for these sources is included here; properties
derived from the cross-field observations are provided in Table~3.  

\item Column~38 gives the logarithm of the minimum false-positive probability
run with {\sc wavdetect} in which each source was detected (see $\S$~3.2).  A
lower false-positive probability indicates a more significant source detection.
Note that \nemseven\ ($\approx$86\%) and \nemeight\ ($\approx$78\%) of our
sources are detected with false-positive probability thresholds of
1~$\times$~10$^{-7}$ and 1~$\times$~10$^{-8}$, respectively.

\item Column~39 gives notes on the sources.
``D'' denotes a source detected in more than one of the four observational
fields.
``U'' denotes objects lying in the \hbox{UDF} (see Figure~2).
``G'' denotes objects that were identified as Galactic stars through the
optical spectrophotometric COMBO-17 survey (Wolf~\etal~2004).
``O'' refers to objects that have large cross-band (i.e., between the three
standard bands) positional offsets ($>2 \arcsec$); all of these sources lie at
off-axis angles of $>8\arcmin$. 
``M'' refers to sources where the photometry was performed manually. 
``S'' refers to close-double or close-triple sources where manual separation
was required.  
``C'' refers to sources detected within the boundary of the $\approx$1~Ms CDF-S
exposure that have no A03 or G02 counterparts.  Several of these sources are
located in low-sensitivity regions of the $\approx$1~Ms CDF-S, and a few of
these sources may be variable.  For further explanation of many
of these notes, see the above text in this section on manual correction of the
{\sc wavdetect} results.

\end{itemize}

In Table~3 we summarize the cross-field source properties of the 14 sources
detected in more than one observational field; none of these sources were
detected in more than two fields.  These properties were derived from the
observation not included in the main \chandra\ catalog (see columns~37 and 39
of Table~2) and are included here for comparison.  The columns of Table~3 are
the same as those in Table~2; the source number for each source corresponds to
its duplicate listed in the main \chandra\ catalog.  In Table~4 we summarize
the source detections in the three standard bands for each of the observational
fields and the main {\it Chandra} catalog.  In total \nsources\ point sources
are detected in one or more of the three standard bands; 14 of these sources
are detected in more than one of the four observational fields (see columns~37
and 39 of Tables~2 and 3) leaving a total of \nunique\ distinct point sources.
Out of these \nunique\ distinct point sources, we find that \ncdfsunique\ are
coincident with sources included in the main {\it Chandra} catalog for the
$\approx$1~Ms CDF-S presented in A03 (see columns~19--24 of Table~2).  For
these sources, we find reasonable agreement between the derived \hbox{X-ray}
properties presented here and in A03.  A total of \nnew\ new point sources are
thus detected here, which brings the total number of E-CDF-S plus $\approx$1~Ms
CDF-S sources to \ntot.  

In Table~5 we summarize the number of sources detected in one band but not
another.  All but two of the detected sources are detected in either the soft
or full bands.  From Tables~4 and 5, the fraction of hard-band sources not
detected in the soft band is $96/453 \approx 21\%$.  The fraction is somewhat
higher than for the \chandra\ Deep Fields, where it is $\approx 14\%$.  Some of
this difference is likely due to differing methods of cross-band matching
(i.e., compare $\S$~3.4.1 of A03 with our $\S$~3.3.1).  Furthermore, this
fraction is physically expected to vary somewhat with sensitivity limit.  We
have also attempted comparisons with \hbox{X-ray} surveys of comparable depth
(Stern et~al. 2002; Harrison et~al. 2003; Wang et~al.  2004a,b; Nandra et~al.
2005) to the \hbox{E-CDF-S}.  Such comparisons are not entirely straightforward
due to varying energy bands utilized, source-selection techniques, and
source-searching methods.  However, the ``hard-band but not soft-band''
fractions for these surveys appear plausibly consistent ($\approx$15--25\%)
with that for the \hbox{E-CDF-S}.

In Figure~9 we show the distributions of detected counts in the three standard
bands.  There are 154 sources with $>$~100 full-band counts, for which basic
spectral analyses are possible; there are eight sources with $>$~1000 full-band
counts.  Figure~10 shows the distribution of effective exposure time for the
three standard bands.  The median effective exposure times for the soft and
hard bands are $\approx$\sbexp~ks and $\approx$\hbexp~ks, respectively. In
Figure~11 we show the distributions of \hbox{X-ray} flux in the three standard
bands.  The \hbox{X-ray} fluxes in this survey span roughly four orders of
magnitude with $\approx$50\% of the sources having soft band fluxes less than
50~$\times$~10$^{-17}$~\flux, a flux regime that few X-ray surveys have probed
with significant areal coverage.

In Figure~12 we show ``postage-stamp'' images from the WFI $R$-band image with
adaptively-smoothed full-band contours overlaid for sources included in the
main {\it Chandra} catalog.  The wide range of \hbox{X-ray} source sizes
observed in these images is largely due to PSF broadening with off-axis angle.
In Figure~13 we plot the positions of sources detected in the main {\it
Chandra} catalog.  Sources that are also included in the A03 CDF-S source
catalog are indicated as open circles, and new \hbox{X-ray} sources detected in
this survey are indicated as filled circles; the circle sizes depend upon the
most significant false-positive probability run with {\sc wavdetect} for which
each source was detected (see column~38 of Table~2).  The majority of the
sources lie in the vicinities of the aim points where the fields are most
sensitive.  In Figure~14 we show the band ratio as a function of full-band
count rate for sources in the main \chandra\ catalog.  This plot shows that the
mean band ratio for sources detected in both the soft and hard bands hardens
for fainter fluxes, a trend observed in other studies (e.g., della Ceca
~\etal~1999; Ueda~\etal~1999; Mushotzky~\etal~2000; B01; Tozzi~\etal~2001;
A03).  This trend is due to the detection of more absorbed AGNs at low flux
levels, and it has been shown that AGNs will dominate the number counts down to
\hbox{0.5--2.0~keV} fluxes of $\approx$1~$\times$~10$^{-17}$ \flux\ (e.g.,
Bauer~\etal~2004).  Figure~15a shows the $R$-band magnitude versus the soft
band flux for sources included in the main \chandra\ catalog.  The approximate
\hbox{X-ray} to $R$-band flux ratios for AGNs and galaxies (e.g.,
Maccaccaro~\etal~1998; Stocke~\etal~1991; Hornschemeier~\etal~2001;
Bauer~\etal~2004) are indicated with dark and light shading, respectively.  The
majority of the sources in this survey appear to be AGNs.  Sixty-one of the
sources were reliably classified as AGNs and seventeen sources have been
identified as Galactic stars (see column~39 of Table~2) in the COMBO-17 survey
(Wolf~\etal~2004).  A significant minority of the sources appear to have
X-ray-to-optical flux ratios characteristic of normal or starburst galaxies. 

\subsubsection{Supplementary Optically Bright \chandra\ Source Catalog}

The density of optically bright ($R < 23$) sources on the sky is comparatively
low. Therefore, we can search for \hbox{X-ray} counterparts to optically
bright sources at a lower \hbox{X-ray} significance threshold than that
used in the main catalog without introducing many false sources
(see \S5.3 of Richards et~al. 1998 for a similar technique applied at
radio wavelengths). We ran {\sc wavdetect} with a false-positive probability
threshold of $1\times 10^{-5}$ on images created in the three
standard bands. A basic lower significance \chandra\ catalog was produced
containing 323 \hbox{X-ray} sources not present in the main \chandra\ source
catalog.

In our matching of these lower significance \chandra\ sources to optically
bright sources, we used the WFI $R$-band source catalog described in \S3.3. We
searched for \hbox{X-ray} counterparts to these optical sources using a
matching radius of $1\farcs 3$. Based upon offset tests, as described below, we
found empirically that we could match to sources as faint as $R=23$ without
introducing an unacceptable number of false matches; this $R$-band cutoff
provides an appropriate balance between the number of detected sources and the
expected number of false sources.

In total 26 optically bright \hbox{X-ray} sources were found via our
matching. We estimated the expected number of false matches by
artificially offsetting the \hbox{X-ray} source coordinates in RA and Dec
by both $5\arcsec$ and $10\arcsec$ (using both positive and negative shifts)
and then re-correlating with the optical sources. On average $\approx$3
matches were found with these tests, demonstrating that the majority
of the 26 \hbox{X-ray} matches are real \hbox{X-ray} sources; only
about 12\% of these sources are expected to be spurious matches.

We also included seven $R<21$ sources where the \hbox{X-ray} source lay
\hbox{$1\farcs 3$--$10\farcs 0$} from the centroid of the optical source but
was still within the extent of the optical emission.  Using optical
spectrophotometric redshift information from COMBO-17, we required that our
off-nuclear sources have \hbox{0.5--2.0~keV} luminosities of $\simlt$$10^{40}$
\xlum.  This restriction was intended to remove obvious sources not associated
with their host galaxies; this led to the removal of one candidate source
(J$033210.9-280230$) when forming our sample of seven plausible off-nuclear
\hbox{X-ray} sources.  Of the seven selected off-nuclear sources included in
the supplementary catalog, we found that six of the host galaxies have COMBO-17
redshift information available. These galaxies were found to have $z \approx
0.10-0.25$ and \hbox{0.5--2.0~keV} luminosities in the range of
$\approx$$10^{39-40}$ \xlum. These derived luminosities are consistent with
these sources being off-nuclear \hbox{X-ray} binaries or star-forming regions
associated with bright host galaxies.  Since these seven sources were
identified in a somewhat subjective manner, it is not meaningful to determine a
false-matching probability for them. These sources are indicated in column 33
of Table~6.  Thus, in total, the supplementary optically bright \chandra\
source catalog contains 33 sources.

The format of Table~6 is similar to that of Table~2.  Details of the columns in
Table~6 are given below.

\begin{itemize}

\item
Column~1 gives the source number (see column~1 of Table~2 for details).

\item Columns~2 and 3 give the RA and Dec of the \hbox{X-ray} source,
respectively. The {\sc wavdetect} positions are given for these faint
\hbox{X-ray} sources. Whenever possible, we quote the position determined in
the full band; when a source is not detected in the full band we use, in order
of priority, the soft-band position and then the hard-band position. The
priority ordering of position choices above was designed to maximize the
signal-to-noise of the data being used for positional determination.

\item Column~4 gives the positional uncertainty in arcseconds.  For these faint
\hbox{X-ray} sources, the positional uncertainty is take to be $1\farcs2$, the
90th percentile of the average optical-X-ray positional offsets given in
column~17.

\item Column~5 gives the off-axis angle for each source in arcminutes (see
column~5 of Table~2 for details).

\item Columns~6--14 give the counts and the corresponding 1$\sigma$ upper and
lower statistical errors (using Gehrels 1986), respectively, for the three
standard bands. The photometry is taken directly from {\sc wavdetect} for these
faint \hbox{X-ray} sources.

\item Columns~15 and 16 give the RA and Dec of the optical source centroid,
respectively.

\item Column~17 gives the measured offset between the optical and \hbox{X-ray}
sources (i.e., $O-X$) in arcseconds.

\item
Column~18 gives the $R$-band magnitude (AB) of the optical source.

\item Column~19 gives the $\approx$1~Ms CDF-S source number from the main
\chandra\ catalog presented in A03 (see column~1 of Table~3a in A03) for
supplementary sources that were matched to A03 counterparts.  We used a
matching radius of 1.5 times the sum of the positional errors of the
\hbox{E-CDF-S} and A03 source positions.  We note that for each matched source
only one match was observed; supplementary sources with no A03 match have a
value of ``0''.

\item Columns~20 and 21  give the RA and Dec of the corresponding $\approx$1~Ms
CDF-S A03 source indicated in column~19.  Sources with no A03 match have RA and
Dec values set to \hbox{``00 00 00.00''} and \hbox{``+00 00 00.0''}.

\item Column~22 gives the $\approx$1~Ms CDF-S source number from the main
\chandra\ catalog presented in G02 (see ``ID'' column of Table~2 in G02) for
supplementary sources that were matched to G02 counterparts.  When matching our
supplementary source positions with G02 counterparts, we removed noted offsets
to the G02 positions of $-1\farcs2$ in RA and $+0\farcs8$ in Dec (see $\S$~A3
of A03); these positions are corrected in the quoted source positions in
columns~23 and 24.  We used a matching radius of 1.5 times the \hbox{E-CDF-S}
positional error plus the G02-quoted positional error for each source position.
We note that for each matched source only one match was observed; supplementary
sources with no G02 match have a value of ``0''.

\item Columns~23 and 24 give the RA and Dec of the corresponding $\approx$1~Ms
CDF-S G02 source indicated in column~22.  Note that the quoted positions have
been corrected by the noted offsets described in column~22 (see $\S$~A3 of
A03). Sources with no G02 match have RA and Dec values set to \hbox{``00 00
00.00''} and \hbox{``+00 00 00.0''}. 

\item Columns~25--27 give the effective exposure times derived from the
standard-band exposure maps (see columns~25--27 of Table~2 for details).

\item Column~28 gives the photon index used to calculate source fluxes
(columns~26--28).  We used a constant photon index of $\Gamma=2.0$ since our
source-selection technique preferentially selects objects with flux-ratios
$f_{\rm 0.5-2.0~keV}/f_R < 0.1$, which are observed to have effective photon
indices of $\Gamma \approx 2$ (e.g., $\S$~4.1.1 of Bauer \etal~2004).

\item Column~29--31 give observed-frame fluxes in the three standard bands;
quoted fluxes are in units of $10^{-15}$ \flux\ and have been calculated assuming
$\Gamma=2.0$.  The fluxes have not been corrected for absorption by the Galaxy
or material intrinsic to the sources (see \hbox{columns~34--36} of Table~2 for
details).

\item Column~32 gives the observational field number corresponding to the
detected source (see column~37 of Table~2 for details).

\item Column~33 gives notes on the sources. With the exception of the
additional note given below, the key for these notes is given in column~39 of
Table~2.
``L'' refers to objects where the \hbox{X-ray} source lies $>1\farcs 3$ from
the centroid of the optical source but is still within the extent of the
optical emission (see the text above for further discussion).

\end{itemize}

The $R$-band magnitudes of the supplementary sources span
\hbox{$R=$~16.6--22.9}.  In Figure~15b we show the $R$-band magnitude versus
soft-band flux.  All of the sources lie in the region expected for starburst
and normal galaxies.  Three of these sources have been classified as Galactic
stars via optical classifications from COMBO-17 (see column~33 of Table~6).
Some of these sources may be low-luminosity AGNs; the small number of hard-band
detections ($\approx$6$\%$) indicates that few of these are absorbed AGNs.  Due
to the low number of counts detected for these sources, we do not provide
postage-stamp images as we did for sources in the main catalog (i.e.,
Figure~12).

The addition of the optically bright supplementary sources increases the number
of extragalactic objects in the \hbox{E-CDF-S} with $f_{\rm 0.5-2.0~keV}/f_R <
0.1$ and $f_{\rm 0.5-2.0~keV}/f_R < 0.01$ by $\approx$20\% and $\approx$50\%,
respectively.  However, the optically bright supplementary sources are not
representative of the faintest \hbox{X-ray} sources as a whole since our
selection criteria preferentially select optically bright and \hbox{X-ray}
faint non-AGNs (e.g., A03; Hornschemeier~\etal~2003).

%
%%%%%%%%%%%%%%%%%%%%%%%%%%%%%%%%%%%%%%%%%%%%%%%%%%%%%%%%%%%%%%%%%%%%%%%%%%%%%%%%%%
\section{Background and Sensitivity Analysis}
%%%%%%%%%%%%%%%%%%%%%%%%%%%%%%%%%%%%%%%%%%%%%%%%%%%%%%%%%%%%%%%%%%%%%%%%%%%%%%%%%%
%

The faintest sources in the main \chandra\ catalog have $\approx$\minsbcnts\
counts (see Table~4).
For a $\Gamma=$~1.4 power law with Galactic absorption, the corresponding
soft-band and hard-band fluxes at the aim points are $\approx$\minsbflux\
erg~cm$^{-2}$~s$^{-1}$ and $\approx$\minhbflux\ erg~cm$^{-2}$~s$^{-1}$,
respectively. This gives a measure of the ultimate sensitivity of this survey,
however, these numbers are only relevant for a small region close to the aim
point. To determine the sensitivity across the field it is necessary
to take into account the broadening of the PSF with off-axis angle, as well as
changes in the effective exposure and background rate across the field.  We
estimated the sensitivity across the field by employing a Poisson model, which
was calibrated for sources detected in the main \chandra\ catalog.  Our
resulting relation is 

\begin{equation}
\log(N)~=\alpha + \beta \log(b) + \gamma [\log(b)]^2 + \delta [\log(b)]^3
\end{equation}

\noindent where $N$ is the required number of source counts for detection, and
$b$ is the number of background counts in a source cell; $\alpha = 0.967$,
$\beta = 0.414$, $\gamma = 0.0822$, and $\delta = 0.0051$ are fitting constants.
The only component within this equation that we need to measure is the
background. For the sensitivity calculations here we measured the background in
a source cell using the background maps described below and assumed an aperture
size of 70\% of the encircled-energy radius of the PSF; the 70\% encircled
energy-radius was chosen as a compromise between having too few source counts
and too many background counts. The total background includes contributions
from the unresolved cosmic background, particle background, and instrumental
background (e.g.,\ Markevitch 2001; Markevitch \etal 2003). For our analyses we
are only interested in the total background and do not distinguish between
these different components.

To create background maps for all of the twelve images, we first masked out the
point sources from the main \chandra\ catalog using apertures with radii twice
that of the $\approx$~90\% PSF encircled-energy radii. The resultant images
should include minimal contributions from detected point sources. They will,
however, include contributions from extended sources (e.g., Bauer \etal 2002;
see $\S$~6), which will cause a slight overestimation of the measured
background close to extended sources.  Extensive testing of background-count
distributions in all three standard bands has shown that the \hbox{X-ray}
background follows a nearly Poisson count distribution (see $\S$~4.2 of A03).
We filled in the masked regions for each source with a local background
estimate by making a probability distribution of counts using an annulus with
inner and outer radii of 2 and 4 times the $\approx$90\% PSF encircled-energy
radius, respectively.  The background properties are summarized in Table~7, and
the full-band background map is shown in Figure~16. The majority of the pixels
have no background counts (e.g., in the full band $\approx$97\% of the pixels
are zero) and the mean background count rates for these observations are
broadly consistent with those presented in A03.  

Following equation 4, we generated sensitivity maps using the background and
exposure maps; we assumed a $\Gamma=$~1.4 power-law model with Galactic
absorption. In Figure~17 we show the full-band sensitivity map, and in
Figure~18 we show plots of flux limit versus solid angle for the full, soft,
and hard bands. The $\approx$1~arcmin$^2$ regions at the aim points has
average \hbox{0.5--2.0~keV} and \hbox{2--8~keV} sensitivity limits of
\hbox{$\approx$\sblimit~erg~cm$^{-2}$~s$^{-1}$} and
\hbox{$\approx$\hblimit~erg~cm$^{-2}$~s$^{-1}$}, respectively. Since we do not filter
out detected sources with our sensitivity maps, a number of sources have fluxes below
these sensitivity limits (4 sources in the soft band and 17 sources in the
hard band). Approximately 800~arcmin$^2$ of the field (i.e.,\ $\approx$~3
times the size of a single ACIS-I field) has a soft-band sensitivity limit of
$\simlt 3\times10^{-16}$~erg~cm$^{-2}$~s$^{-1}$, well into the flux range where
few X-ray surveys have probed (see Figure~1). 

%
%%%%%%%%%%%%%%%%%%%%%%%%%%%%%%%%%%%%%%%%%%%%%%%%%%%%%%%%%%%%%%%%%%%%%%%%%%%%%%%%%%
\section{Number Counts for Main \chandra\ Catalog}
%%%%%%%%%%%%%%%%%%%%%%%%%%%%%%%%%%%%%%%%%%%%%%%%%%%%%%%%%%%%%%%%%%%%%%%%%%%%%%%%%%
%

We have calculated cumulative number counts, $N(>S)$, for the soft and hard
bands using sources presented in our main \chandra\ catalog (see Table~2).   We
restricted our analyses to flux levels where we expect to be mostly complete
based on our sensitivity maps and simulations performed by Bauer et~al. (2004);
this also helps to guard against Eddington bias at low flux levels. We
empirically set our minimum flux levels to 3.0~$\times$~10$^{-16}$ \flux\ in
the soft band and 1.2~$\times$~10$^{-15}$ \flux\ in the hard band, which
correspond to the minimum detected fluxes for sources with $\simgt$15~counts in
each respective band.

Assuming completeness to the flux levels quoted above, the cumulative number of
sources, $N(>S)$, brighter than a given flux, $S$, weighted by the appropriate
aerial coverage, is

\begin{equation}
N(>S) = \sum_{S_i > S} \Omega_i^{-1}
\end{equation}

\noindent where $\Omega_i$ is the maximum solid angle for which a source with
measured flux, $S_i$, could be detected.  Each maximum solid angle was computed
using the profiles presented in Figure~18.  In Figure~19 we show the cumulative
number counts for the main \chandra\ catalog.  Number counts derived for the
$\approx$1~Ms \hbox{CDF-S} from Rosati et~al. (2002) have been plotted for
comparison.  The \hbox{E-CDF-S} number counts appear to be consistent with
those from the $\approx$1~Ms \hbox{CDF-S} to within $\approx$1~$\sigma$ over
the overlapping flux ranges.  We note, however, that the hard-band number
counts appear to be generally elevated with respect to those from the
$\approx$1~Ms \hbox{CDF-S}; this effect is likely a signature of field-to-field
variance from the $\approx$1~Ms \hbox{CDF-S} where a smaller solid angle is
surveyed.  Even with the conservative flux constraints used in our
number-counts analysis we reach source densities exceeding $\approx$2000
deg$^{-2}$; as noted in $\S$~3.3.1, a large majority of these sources are AGNs.
For comparison, the number density of COMBO-17 sources with reliable AGN
identifications is $\approx$300 deg$^{-2}$ (Wolf et~al. 2004).

%%%%%%%%%%%%%%%%%%%%%%%%%%%%%%%%%%%%%%%%%%%%%%%%%%%%%%%%%%%%%%%%%%%%%%%%%%%%%%%%%%
\section{Extended Sources}
%%%%%%%%%%%%%%%%%%%%%%%%%%%%%%%%%%%%%%%%%%%%%%%%%%%%%%%%%%%%%%%%%%%%%%%%%%%%%%%%%%
%

We searched the standard-band images for extended sources using the Voronoi
Tessellation and Percolation algorithm {\sc vtpdetect}
(Ebeling~\&~Wiedenmann~1993; Dobrzycki et~al. 2002).  In our {\sc vtpdetect}
searching, we adopted a false-positive probability threshold of
1~$\times$~10$^{-7}$ and a ``coarse'' parameter of 50.  Following the
source-detection criteria presented in Bauer~\etal~(2002), we further required
that {\sc vtpdetect}-detected sources have (1) average {\sc vtpdetect} radii
(i.e., average of the 3$\sigma$ major and minor axes estimated by {\sc
vtpdetect}) $\ge$~three times the 95\% encircled-energy radius of a point
source at the given position and (2) visible evidence for extended emission in
the adaptively smoothed, exposure-corrected images (see $\S$~3.3.1 and
Figure~4).  Application of these somewhat conservative selection criteria
yielded three extended X-ray sources, all of which are detected only in the
soft band.  The soft emission from the most significant of these three extended
sources, CXOECDFS \hbox{J033320.3$-$274836}, is clearly visible as an extended
red ``glow'' near the left-hand side of Figure~7.

The \hbox{X-ray} properties of these three extended sources are presented in
Table~8; our analysis was limited to the soft band where we find all of our
detections.  The counts for extended sources were determined using manual
aperture photometry; point sources were masked out using circular apertures
with radii of twice the 95\% encircled-energy radii (see Footnote~3).  We
extracted extended-source counts using elliptical apertures with sizes and
orientations that most closely matched the apparent extent of X-ray emission
($>$~10\% above the background level) as observed in the adaptively
smoothed images (see Table~8).  The local background was estimated using
elliptical annuli with inner and outer sizes of 1 and 2.5 times those used for
extracting source counts.  In order to calculate properly the expected numbers
of background counts in our source extraction regions, we extracted total
exposure times from the source and background regions (with point sources
removed) and normalized the extracted background counts to the source exposure
times.  That is, using the number of background counts $b_{\rm m}$ and total
background exposure time $T_{\rm m}$ as measured from the elliptical annuli, we
calculated the expected number of background counts $b_{\rm s}$ in a source
extraction region with total exposure time $T_{\rm s}$ as being $b_{\rm s} =
b_{\rm m} T_{\rm s}/T_{\rm m}$.  This technique was used to account
for extended emission from sources that are spatially distributed over more
than one observational field, which was the case for CXOECDFS
\hbox{J033320.3$-$274836}.

Figure~20 shows WFI $R$-band images of the extended sources with adaptively
smoothed soft-band contours overlaid.  Inspection of the spectrophotometric
redshifts of optical sources (from COMBO-17) in these regions suggests that the
extended \hbox{X-ray} emission for all three sources may originate from
low-to-moderate redshift groups or poor clusters.  The most conspicuous of these
is the apparent clustering of galaxies at $z \approx 0.1$ in the area of
CXOECDFS \hbox{J033320.3$-$274836}, an $\approx$20~arcmin$^2$ extended
\hbox{X-ray} source.  CXOECDFS \hbox{J033209.6$-$274242} lies in the
$\approx$1~Ms CDF-S and was previously detected as an extended source by G02.
Optical spectroscopic follow-up observations using the VLT have shown that this
source is associated with a galaxy cluster at \hbox{$z = 0.73$} (Szokoly et~al.
2004).  Suggestive evidence for clustering at $z \approx 0.7$ is also observed
for CXOECDFS \hbox{J033257.9$-$280155}; this may be an extension of the
large-scale structures observed in the $\approx$1~Ms CDF-S (Gilli~\etal~2003, 2005).
Under the assumption that these sources are indeed groups or poor clusters at the
discussed redshifts, we computed the expected soft-band fluxes and luminosities
assuming a Raymond-Smith thermal plasma (Raymond \& Smith 1977) with \hbox{$kT
= 1.0$~keV} (see Table~8).  We find that these sources would have rest-frame
{0.5--2.0~keV} luminosities of $\approx$1--5~$\times 10^{42}$~\xlum.  Further
optical spectroscopic observations and analyses beyond the scope of this paper
are required for confirmation of the nature of these sources.

%
%%%%%%%%%%%%%%%%%%%%%%%%%%%%%%%%%%%%%%%%%%%%%%%%%%%%%%%%%%%%%%%%%%%%%%%%%%%%%%%%%%
\section{Summary}
%%%%%%%%%%%%%%%%%%%%%%%%%%%%%%%%%%%%%%%%%%%%%%%%%%%%%%%%%%%%%%%%%%%%%%%%%%%%%%%%%%
%

We have presented catalogs and basic analyses of point sources detected in the
250~ks $\approx$0.3 deg$^2$ Extended {\it Chandra} Deep Field-South
(\hbox{E-CDF-S}).  The survey area consists of four observational fields, of
similar exposure, with average on-axis flux limits in the \hbox{0.5--2.0~keV}
and \hbox{2--8~keV} bandpasses of
\hbox{$\approx$\sblimit~erg~cm$^{-2}$~s$^{-1}$} and
\hbox{$\approx$\hblimit~erg~cm$^{-2}$~s$^{-1}$}, respectively.  We have
presented two catalogs: a main \chandra\ catalog of \nunique\ sources (\nnew\
of these are new), which was generated by running {\sc wavdetect} with a
false-positive probability threshold of 1~$\times$~10$^{-6}$, and a
supplementary catalog of 33 lower-significance (false-positive probability
threshold of 1~$\times$~10$^{-5}$) \hbox{X-ray} sources with optically bright
($R$~$<$~23) counterparts.  The \hbox{X-ray} spectral properties and optical
fluxes of sources in our main \chandra\ catalog indicate a variety of source
types, most of which are absorbed AGNs that dominate at lower \hbox{X-ray}
fluxes.  The \hbox{X-ray} and optical properties of sources in the
supplementary optically bright \chandra\ catalog are mostly consistent with
those expected for starburst and normal galaxies.  We have presented basic
number-count results for point sources in our main \chandra\ catalog and find
overall consistency with number counts derived for the $\approx$1~Ms
\hbox{CDF-S} in both the \hbox{0.5--2.0~keV} and \hbox{2--8~keV} bandpasses.
We have also presented three \hbox{0.5--2.0~keV} extended sources, which were
detected using a conservative detection criterion.  These sources are likely
associated with groups or poor clusters at $z \approx 0.1-0.7$ with
\Lx~$\approx$~1--5 $\times 10^{42}$~\xlum.

\acknowledgements

We thank G. Chartas, P.E. Freeman, A.T. Steffen, and L.K. Townsley for helpful
discussions.  We thank the anonymous referee for useful comments that improved
the manuscript. 
Support for this work was provided by NASA through \chandra\ Award Number
GO4-5157 (BDL, WNB, DPS, RG, AMK, TM) issued by the \chandra\ \hbox{X-ray}
Observatory Center, which is operated by the Smithsonian Astrophysical
Observatory under NASA contract NAS8-03060.  We also acknowledge the financial
support of NSF CAREER award AST-9983783 (BDL, WNB), the Royal Society (DMA),
the Chandra Fellowship program (FEB), NSF AST 03-07582 (DPS), NASA LTSA Grant
NAG5-10875 (TM), and MIUR COFIN grant 03-02-23 (CV).

\clearpage

%
%%%%%%%%%%%%%%%%%%%%%%%%%%%%%%%%%%%%%%%%%%%%%%%%%%%%%%%%%%%%%%%%%%%%%%%%%%%%%%%%%%

%\clearpage

%
%%%%%%%%%%%%%%%%%%%%%%%%%%%%%%%%%%%%%%%%%%%%%%%%%%%%%%%%%%%%%%%%%%%%%%
% TABLE 1
%%%%%%%%%%%%%%%%%%%%%%%%%%%%%%%%%%%%%%%%%%%%%%%%%%%%%%%%%%%%%%%%%%%%%%
%
\tabletypesize{\small}
\begin{deluxetable}{lccccccc}
\tablenum{1}
% \tabletypesize{\footnotesize}
\tablecaption{Journal of Extended {\it Chandra} Deep Field-South Observations}
\tablehead{
\colhead{Obs.}                                 &
\colhead{Obs.}                                 &
\colhead{Exposure}                             &
\multicolumn{2}{c}{Aim Point}                  &
\colhead{Roll}                                 &
\colhead{Field}                                &
\colhead{Pipeline}                             \\
\colhead{ID}                                 &
\colhead{Start (UT)}                         &
\colhead{Time (ks)$^{\rm a}$ }               &
\colhead{$\alpha_{\rm 2000}$}                &
\colhead{$\delta_{\rm 2000}$}                &
\colhead{Angle ($^\circ$)$^{\rm b}$}         &
\colhead{Number}                             &
\colhead{Version}                             
}
\tablewidth{0pt}
\startdata
5015\dotfill \ldots \ldots & 2004-02-29, 21:21 & 162.9 & 03 33 05.61 & $-$27 41 08.84 & 270.2 & 1 & 7.1.1 \\
5016\dotfill \ldots \ldots & 2004-03-03, 12:09 &  77.2 & 03 33 05.61 & $-$27 41 08.84 & 270.2 & 1 & 7.2.0 \\
5017\dotfill \ldots \ldots & 2004-05-14, 01:09 & 155.4 & 03 31 51.43 & $-$27 41 38.80 & 181.5 & 2 & 7.2.1 \\
5018\dotfill \ldots \ldots & 2004-05-16, 13:44 &  72.0 & 03 31 51.43 & $-$27 41 38.80 & 181.5 & 2 & 7.2.1 \\
5019\dotfill \ldots \ldots & 2004-11-17, 14:43 & 163.1 & 03 31 49.94 & $-$27 57 14.56 &   0.2 & 3 & 7.3.2 \\
5020\dotfill \ldots \ldots & 2004-11-15, 23:25 &  77.6 & 03 31 49.94 & $-$27 57 14.56 &   0.2 & 3 & 7.3.2 \\
5021\dotfill \ldots \ldots & 2004-11-13, 03:26 &  97.8 & 03 33 02.93 & $-$27 57 16.08 &   0.2 & 4 & 7.3.2 \\
5022\dotfill \ldots \ldots & 2004-11-15, 00:51 &  79.1 & 03 33 02.93 & $-$27 57 16.08 &   0.2 & 4 & 7.3.2 \\
6164\dotfill \ldots \ldots & 2004-11-20, 21:08 &  69.1 & 03 33 02.93 & $-$27 57 16.08 &   0.2 & 4 & 7.3.2 \\
\enddata

\tablenotetext{a}{All observations were continuous.  The short time
intervals with bad satellite aspect are negligible and have not been removed.}

\tablenotetext{b}{Roll angle describes the orientation of the \chandra\ instruments
on the sky (see Figure~2).  The angle is between 0--360$^{\circ}$, and it increases to the West
of North (opposite to the sense of traditional position angle).}

\end{deluxetable}

%
%%%%%%%%%%%%%%%%%%%%%%%%%%%%%%%%%%%%%%%%%%%%%%%%%%%%%%%%%%%%%%%%%%%%%%%%%%%%%%%%%%%
% Table 2
%%%%%%%%%%%%%%%%%%%%%%%%%%%%%%%%%%%%%%%%%%%%%%%%%%%%%%%%%%%%%%%%%%%%%%%%%%%%%%%%%%%
%

\begin{deluxetable}{lllcccccccc}
\tablenum{2}
\tabletypesize{\scriptsize}
\tablewidth{0pt}
\tablecaption{Main {\it Chandra} Catalog Source Properties}

\tablehead{
\colhead{} &
\multicolumn{2}{c}{X-ray Coordinates} &
\colhead{}                   &
\colhead{}                   &
\multicolumn{6}{c}{Counts}      \\

\\ \cline{2-3} \cline{6-11}  \\

\colhead{No.}                    &
\colhead{$\alpha_{2000}$}       &
\colhead{$\delta_{2000}$}       &
\colhead{Pos Err}       &
\colhead{Off-Axis}       &
\colhead{FB}          &
\colhead{FB Upp Err}          &
\colhead{FB Low Err}          &
\colhead{SB}          &
\colhead{SB Upp Err}          &
\colhead{SB Low Err}          \\

\colhead{(1)}         &
\colhead{(2)}         &
\colhead{(3)}         &
\colhead{(4)}         &
\colhead{(5)}         &
\colhead{(6)}         &
\colhead{(7)}         &
\colhead{(9)}         &
\colhead{(10)}        &
\colhead{(11)}        &
\colhead{(12)}         
}

\startdata
       1 \dotfill \ldots   &03 31 11.40 &$-$27 33 38.5& 2.6&11.95&     29.0&    11.7&    10.6&     20.2&    $-$1&    $-$1\\
       2 \dotfill \ldots   &03 31 12.99 &$-$27 55 48.2& 0.8& 8.29&    326.8&    20.9&    19.8&    152.1&    14.2&    13.2\\
       3 \dotfill \ldots   &03 31 14.09 &$-$28 04 20.3& 2.3&10.63&     27.1&    $-$1&    $-$1&     21.3&     7.6&     6.4\\
       4 \dotfill \ldots   &03 31 14.24 &$-$27 47 07.3& 0.8& 9.87&     99.2&    13.4&    12.3&     76.5&    10.7&     9.7\\
       5 \dotfill \ldots   &03 31 14.64 &$-$28 01 44.3& 0.8& 9.00&     73.3&    11.9&    10.8&     27.4&     7.5&     6.3\\
\enddata

\tablecomments{Table~2 is presented in its entirety in the electronic edition.  An abbreviated version of the table is shown here for guidance as to its form and content.  The full table contains 39 columns of information on the 762 \hbox{X-ray} sources.}

\end{deluxetable}

%
%%%%%%%%%%%%%%%%%%%%%%%%%%%%%%%%%%%%%%%%%%%%%%%%%%%%%%%%%%%%%%%%%%%%%%%%%%%%%%%%%%%
% Table 3
%%%%%%%%%%%%%%%%%%%%%%%%%%%%%%%%%%%%%%%%%%%%%%%%%%%%%%%%%%%%%%%%%%%%%%%%%%%%%%%%%%%
%

\begin{deluxetable}{lllcccccccc}
\tablenum{3}
\tabletypesize{\scriptsize}
\tablewidth{0pt}
\tablecaption{Main {\it Chandra} Catalog Cross-Field Source Properties}

\tablehead{
\colhead{} &
\multicolumn{2}{c}{X-ray Coordinates} &
\colhead{}                   &
\colhead{}                   &
\multicolumn{6}{c}{Counts}      \\

\\ \cline{2-3} \cline{6-11}  \\

\colhead{No.}                    &
\colhead{$\alpha_{2000}$}       &
\colhead{$\delta_{2000}$}       &
\colhead{Pos Err}       &
\colhead{Off-Axis}       &
\colhead{FB}          &
\colhead{FB Upp Err}          &
\colhead{FB Low Err}          &
\colhead{SB}          &
\colhead{SB Upp Err}          &
\colhead{SB Low Err}          \\

\colhead{(1)}         &
\colhead{(2)}         &
\colhead{(3)}         &
\colhead{(4)}         &
\colhead{(5)}         &
\colhead{(6)}         &
\colhead{(7)}         &
\colhead{(8)}         &
\colhead{(9)}         &
\colhead{(10)}        &
\colhead{(11)}         
}

\startdata
     367 \dotfill \ldots   &03 32 25.22 &$-$27 42 19.4& 0.8& 9.02&     62.7&    11.2&    10.1&     38.0&     8.1&     7.1\\
     369 \dotfill \ldots   &03 32 25.62 &$-$27 43 05.8& 0.8& 9.06&     81.1&    12.1&    11.0&     53.4&     9.2&     8.1\\
     372 \dotfill \ldots   &03 32 25.91 &$-$28 00 46.7& 1.8& 8.90&     22.3&     8.7&     7.6&     14.4&    $-$1&    $-$1\\
     373 \dotfill \ldots   &03 32 26.18 &$-$27 37 12.1& 2.0& 9.58&     35.1&    10.3&     9.2&     22.0&     7.3&     6.1\\
     375 \dotfill \ldots   &03 32 26.49 &$-$27 40 35.4& 0.7& 7.84&    295.8&    20.1&    18.9&    244.0&    17.7&    16.6\\
     376 \dotfill \ldots   &03 32 26.52 &$-$27 35 02.4& 0.9&10.21&     95.3&    13.8&    12.7&     72.8&    10.9&     9.9\\
     377 \dotfill \ldots   &03 32 26.65 &$-$27 40 14.0& 0.8& 8.67&    185.9&    17.1&    16.0&    110.1&    12.9&    11.8\\
     379 \dotfill \ldots   &03 32 27.00 &$-$27 41 05.2& 0.7& 7.90&   1755.3&    44.0&    43.0&   1175.1&    36.3&    35.3\\
     381 \dotfill \ldots   &03 32 27.12 &$-$28 01 24.4& 0.8& 9.21&    425.9&    23.9&    22.7&    347.7&    20.5&    19.5\\
     382 \dotfill \ldots   &03 32 27.37 &$-$28 05 51.6& 0.9&11.94&    865.2&    33.8&    32.7&    622.5&    27.2&    26.1\\
     383 \dotfill \ldots   &03 32 27.62 &$-$27 41 44.9& 0.8& 8.02&     63.6&    11.0&     9.8&     20.3&     6.8&     5.7\\
     387 \dotfill \ldots   &03 32 29.01 &$-$27 57 30.3& 0.8& 8.63&     72.2&    11.2&    10.1&     55.4&     9.1&     8.0\\
     388 \dotfill \ldots   &03 32 29.25 &$-$28 01 46.0& 2.0& 9.79&     12.0&     9.2&     8.1&     12.3&    $-$1&    $-$1\\
     753 \dotfill \ldots   &03 33 38.54 &$-$27 49 42.3& 2.4&11.24&     28.7&    11.3&    10.2&     19.9&    $-$1&    $-$1\\
\enddata

\tablecomments{Basic source properties for the 14 objects detected in more than one observational field.  These properties are derived from the observation not included in the main \chandra\ catalog.  Columns follow the same format as those presented in Table~2; all 39 columns of this table are available via the electronic version.}
\end{deluxetable}

%
%%%%%%%%%%%%%%%%%%%%%%%%%%%%%%%%%%%%%%%%%%%%%%%%%%%%%%%%%%%%%%%%%%%%%%%%%%%%%%%%%%%
% Table 4
%%%%%%%%%%%%%%%%%%%%%%%%%%%%%%%%%%%%%%%%%%%%%%%%%%%%%%%%%%%%%%%%%%%%%%%%%%%%%%%%%%%
%

\begin{deluxetable}{lcccccc}

\tabletypesize{\small}
\tablenum{4}
\tablewidth{0pt}
\tablecaption{Summary of {\it Chandra} Source Detections}

\tablehead{

\colhead{} &
\colhead{} &
\colhead{} &
\multicolumn{4}{c}{Detected Counts Per Source} \\

\colhead{Band (keV)} &
\colhead{Observational Field} &
\colhead{Number of Sources} &
\colhead{Maximum} &
\colhead{Minimum} &
\colhead{Median} &
\colhead{Mean} 
}

\startdata
Full (0.5--8.0) \dotfill \ldots \ldots & 1 & 191 & 2509.4 & 3.9 & 43.4 & 123.2 \\
Soft (0.5--2.0) \dotfill \ldots \ldots & 1 & 167 & 1715.7 & 3.7 & 28.4 & 91.3 \\
Hard (2--8) \dotfill \ldots \ldots & 1 & 128 & 787.7 & 3.9 & 29.0 & 60.8 \\
\\
Full (0.5--8.0) \dotfill \ldots \ldots & 2 & 165 & 2346.5 & 5.2 & 37.7 & 110.2 \\
Soft (0.5--2.0) \dotfill \ldots \ldots & 2 & 144 & 1689.4 & 4.4 & 25.5 & 77.4 \\
Hard (2--8) \dotfill \ldots \ldots & 2 & 109 & 735.2 & 4.3 & 25.2 & 59.2 \\
\\
Full (0.5--8.0) \dotfill \ldots \ldots & 3 & 187 & 1078.3 & 6.0 & 39.3 & 85.4 \\
Soft (0.5--2.0) \dotfill \ldots \ldots & 3 & 156 & 756.6 & 4.6 & 24.3 & 61.1 \\
Hard (2--8) \dotfill \ldots \ldots & 3 & 128 & 324.5 & 3.7 & 26.1 & 45.8 \\
\\
Full (0.5--8.0) \dotfill \ldots \ldots & 4 & 160 & 1771.6 & 4.2 & 37.1 & 115.7 \\
Soft (0.5--2.0) \dotfill \ldots \ldots & 4 & 142 & 1312.0 & 4.2 & 25.8 & 85.0 \\
Hard (2--8) \dotfill \ldots \ldots & 4 & 98 & 514.8 & 4.7 & 28.2 & 60.9 \\
\\
Full (0.5--8.0) \dotfill \ldots \ldots & 1,2,3,4 & 689 & 2509.4 & 3.9 & 38.6 & 104.8 \\
Soft (0.5--2.0) \dotfill \ldots \ldots & 1,2,3,4 & 598 & 1715.7 & 3.7 & 25.9 & 75.7 \\
Hard (2--8) \dotfill \ldots \ldots & 1,2,3,4 & 453 & 787.7 & 3.7 & 27.8 & 55.0 \\
\enddata

\end{deluxetable}

%
%%%%%%%%%%%%%%%%%%%%%%%%%%%%%%%%%%%%%%%%%%%%%%%%%%%%%%%%%%%%%%%%%%%%%%%%%%%%%%%%%%%
% Table 5
%%%%%%%%%%%%%%%%%%%%%%%%%%%%%%%%%%%%%%%%%%%%%%%%%%%%%%%%%%%%%%%%%%%%%%%%%%%%%%%%%%%
%

\begin{deluxetable}{lccc}

\tabletypesize{\small}
\tablenum{5}
\tablewidth{0pt}
\tablecaption{Sources Detected In One Band But Not Another}

\tablehead{

\colhead{} &
\multicolumn{3}{c}{Nondetection Energy Band} \\

\colhead{Detection Band (keV)} &
\colhead{Full} &
\colhead{Soft} &
\colhead{Hard} \\
}

\startdata
Full (0.5--8.0) \dotfill \ldots \ldots & \ldots & 149 & 251 \\
Soft (0.5--2.0) \dotfill \ldots \ldots & 58 & \ldots & 241 \\
Hard (2--8) \dotfill \ldots \ldots & 15 & 96 & \ldots \\
\enddata

\end{deluxetable}

%
%%%%%%%%%%%%%%%%%%%%%%%%%%%%%%%%%%%%%%%%%%%%%%%%%%%%%%%%%%%%%%%%%%%%%%%%%%%%%%%%%%%
% Table 6
%%%%%%%%%%%%%%%%%%%%%%%%%%%%%%%%%%%%%%%%%%%%%%%%%%%%%%%%%%%%%%%%%%%%%%%%%%%%%%%%%%%
%

\begin{deluxetable}{lllcccccccc}
\tablenum{6}
\tabletypesize{\scriptsize}
\tablewidth{0pt}
\tablecaption{Supplementary Optically Bright {\it Chandra} Catalog}

\tablehead{
\colhead{} &
\multicolumn{2}{c}{X-ray Coordinates} &
\colhead{}                   &
\colhead{}                   &
\multicolumn{6}{c}{Counts}      \\

\\ \cline{2-3} \cline{6-11}  \\

\colhead{No.}                    &
\colhead{$\alpha_{2000}$}       &
\colhead{$\delta_{2000}$}       &
\colhead{Pos Err}       &
\colhead{Off-Axis}       &
\colhead{FB}          &
\colhead{FB Upp Err}          &
\colhead{FB Low Err}          &
\colhead{SB}          &
\colhead{SB Upp Err}          &
\colhead{SB Low Err}          \\

\colhead{(1)}         &
\colhead{(2)}         &
\colhead{(3)}         &
\colhead{(4)}         &
\colhead{(5)}         &
\colhead{(6)}         &
\colhead{(7)}         &
\colhead{(8)}         &
\colhead{(9)}         &
\colhead{(10)}         &
\colhead{(11)}         
}

\startdata
         1 \dotfill \ldots &03 31 16.20&$-$27 50 30.9& 1.2 & 10.04&                                  17.9 &      5.3 &      4.2&                                          12.3 & $-$1 & $-1$\\
         2 \dotfill \ldots &03 31 22.00&$-$27 36 20.1& 1.2 &  8.41&                                  17.0 &      5.2 &      4.1&                                          11.5 & $-$1 & $-1$\\
         3 \dotfill \ldots &03 31 28.87&$-$27 53 29.9& 1.2 &  5.97&                                  12.1 &      4.6 &      3.4&                                           9.1 & $-$1 & $-1$\\
         4 \dotfill \ldots &03 31 35.14&$-$27 58 08.6& 1.2 &  3.39&                                          10.0 & $-$1 & $-1$&                                   2.8 &      2.9 &      1.6\\
         5 \dotfill \ldots &03 31 39.05&$-$28 02 21.1& 1.2 &  5.65&                                          14.6 & $-$1 & $-1$&                                   1.5 &      2.5 &      1.1\\
\enddata
\tablecomments{Table~6 is presented in its entirety in the electronic edition.  An abbreviated version of the table is shown here for guidance as to its form and content.  The full table contains 33 columns of information on the 33 \hbox{X-ray} sources.}
\end{deluxetable}

%
%%%%%%%%%%%%%%%%%%%%%%%%%%%%%%%%%%%%%%%%%%%%%%%%%%%%%%%%%%%%%%%%%%%%%%%%%%%%%%%%%%%
% Table 7
%%%%%%%%%%%%%%%%%%%%%%%%%%%%%%%%%%%%%%%%%%%%%%%%%%%%%%%%%%%%%%%%%%%%%%%%%%%%%%%%%%%
%

\begin{deluxetable}{lccccc}

\tabletypesize{\scriptsize}
\tablenum{7}
\tablewidth{0pt}
\tablecaption{Background Parameters}

\tablehead{

\colhead{} &
\colhead{} &
\multicolumn{2}{c}{Mean Background} &
\colhead{Total Background$^{\rm c}$} &
\colhead{Count Ratio$^{\rm d}$} \\

\colhead{Band (keV)} &
\colhead{Observational Field} &
\colhead{(counts pixel$^{-1}$)$^{\rm a}$} &
\colhead{(counts Ms$^{-1}$ pixel$^{-1}$ )$^{\rm b}$} &
\colhead{(10$^4$ counts)} &
\colhead{(background/source)} 
}

\startdata
Full (0.5--8.0)\dotfill \ldots \ldots & 1 & 0.033 & 0.169 & 14.6 &  6.0\\
Soft (0.5--2.0)\dotfill \ldots \ldots & 1 & 0.009 & 0.044 &  3.9 &  2.5\\
    Hard (2--8)\dotfill \ldots \ldots & 1 & 0.024 & 0.128 & 10.8 & 12.7\\
\\
Full (0.5--8.0)\dotfill \ldots \ldots & 2 & 0.036 & 0.192 & 15.8 &  8.6\\
Soft (0.5--2.0)\dotfill \ldots \ldots & 2 & 0.009 & 0.066 &  4.1 &  3.6\\
    Hard (2--8)\dotfill \ldots \ldots & 2 & 0.027 & 0.143 & 11.7 & 16.9\\
\\
Full (0.5--8.0)\dotfill \ldots \ldots & 3 & 0.037 & 0.181 & 16.1 & 10.1\\
Soft (0.5--2.0)\dotfill \ldots \ldots & 3 & 0.009 & 0.046 &  4.0 &  4.2\\
    Hard (2--8)\dotfill \ldots \ldots & 3 & 0.027 & 0.140 & 12.1 & 19.5\\
\\
Full (0.5--8.0)\dotfill \ldots \ldots & 4 & 0.039 & 0.228 & 17.4 &  9.4\\
Soft (0.5--2.0)\dotfill \ldots \ldots & 4 & 0.010 & 0.048 &  4.4 &  3.6\\
    Hard (2--8)\dotfill \ldots \ldots & 4 & 0.030 & 0.148 & 13.1 & 20.6\\
\\
Full (0.5--8.0)\dotfill \ldots \ldots & 1,2,3,4 & 0.036 & 0.192 & 63.9 &  8.5\\
Soft (0.5--2.0)\dotfill \ldots \ldots & 1,2,3,4 & 0.009 & 0.051 & 16.3 &  3.5\\
    Hard (2--8)\dotfill \ldots \ldots & 1,2,3,4 & 0.027 & 0.140 & 47.7 & 17.4\\
\enddata

\tablenotetext{a}{The mean numbers of counts per pixel.  These are measured from the masked background images described in $\S$~4.}
\tablenotetext{b}{The mean numbers of counts per pixel divided by the mean effective exposure.  These are measured from the exposure maps and masked background images described in $\S$~4.}
\tablenotetext{c}{Total number of background counts.}
\tablenotetext{d}{Ratio of the total number of background counts to the total number of source counts.}
\end{deluxetable}

%
%%%%%%%%%%%%%%%%%%%%%%%%%%%%%%%%%%%%%%%%%%%%%%%%%%%%%%%%%%%%%%%%%%%%%%%%%%%%%%%%%%%
% Table 8
%%%%%%%%%%%%%%%%%%%%%%%%%%%%%%%%%%%%%%%%%%%%%%%%%%%%%%%%%%%%%%%%%%%%%%%%%%%%%%%%%%%
%

\begin{deluxetable}{lllccccccc}
\tablenum{8}
\tabletypesize{\scriptsize}
\tablewidth{0pt}
\tablecaption{Extended-Source Properties}

\tablehead{
\colhead{} &
\multicolumn{2}{c}{X-ray Coordinates} &
\colhead{} &
\colhead{} &
\colhead{} &
\colhead{} &
\colhead{} &
\colhead{} &
\colhead{} \\
\colhead{No.} &
\colhead{$\alpha_{2000}$} &
\colhead{$\delta_{2000}$} &
\colhead{Region$^{\rm a}$} &
\colhead{Angle$^{\rm b}$} &
\colhead{SB Counts$^{\rm c}$} &
\colhead{$S$-to-$B$ Ratio$^{\rm d}$} &
\colhead{$z^{\rm e}$} & 
\colhead{SB Flux$^{\rm f}$} &
\colhead{$L_{\mbox{\tiny{X}}}^{\rm g}$}
}

\startdata
   1 \dotfill \ldots \ldots  &03 32 09.62  & $-$27 42 42.2 & 100$\arcsec$ $\times$  60$\arcsec$ & 335$^\circ$ &  44.1 $\pm$  16.5 & 0.22 &  0.7 &  2.2 &  4.8 \\
   2 \dotfill \ldots \ldots  &03 32 57.94  & $-$28 01 55.4 &  90$\arcsec$ $\times$  60$\arcsec$ &  35$^\circ$ &  50.5 $\pm$  16.3 & 0.28 &  0.7 &  1.7 &  4.1 \\
   3 \dotfill \ldots \ldots  &03 33 20.32  & $-$27 48 36.2 & 380$\arcsec$ $\times$ 230$\arcsec$ &  10$^\circ$ & 901.0 $\pm$  60.4 & 0.34 &  0.1 & 27.1 &  0.7 \\
\enddata

\tablenotetext{a}{Extraction region given as major axis and minor axis in arcseconds.}
\tablenotetext{b}{Position angle of the extraction region.}
\tablenotetext{c}{Net 0.5--2.0~keV background-subtracted source counts.  These counts have been measured using the specified extraction regions.}
\tablenotetext{d}{Ratio of the \hbox{0.5--2.0~keV} source counts to the total number of expected \hbox{0.5--2.0~keV} background counts.}
\tablenotetext{e}{Redshift of candidate group or poor cluster associated with the extended source.  All redshifts were inferred from galaxies with optical spectrophotometric redshifts from the COMBO-17 survey with the exception of source number 1, which was previously identified spectroscopically by Szokoly et~al. (2004).}
\tablenotetext{f}{Integrated \hbox{0.5--2.0~keV} X-ray flux in units of $10^{-15}$~\flux, derived for each source assuming a Raymond-Smith thermal plasma spectral energy distribution with $kT = 1.0$~keV at the given redshift.}
\tablenotetext{g}{Integrated \hbox{0.5--2.0~keV} X-ray luminosity in units of $10^{42}$~\xlum.}

\end{deluxetable}

\clearpage

%
%%%%%%%%%%%%%%%%%%%%%%%%%%%%%%%%%%%%%%%%%%%%%%%%%%%%%%%%%%%%%%%%%%%%%%%%%%%%%%%%%%%
% Figure 1
%%%%%%%%%%%%%%%%%%%%%%%%%%%%%%%%%%%%%%%%%%%%%%%%%%%%%%%%%%%%%%%%%%%%%%%%%%%%%%%%%%
%

\begin{figure}
\figurenum{1}
\centerline{
\includegraphics[width=12.0cm,angle=-90]{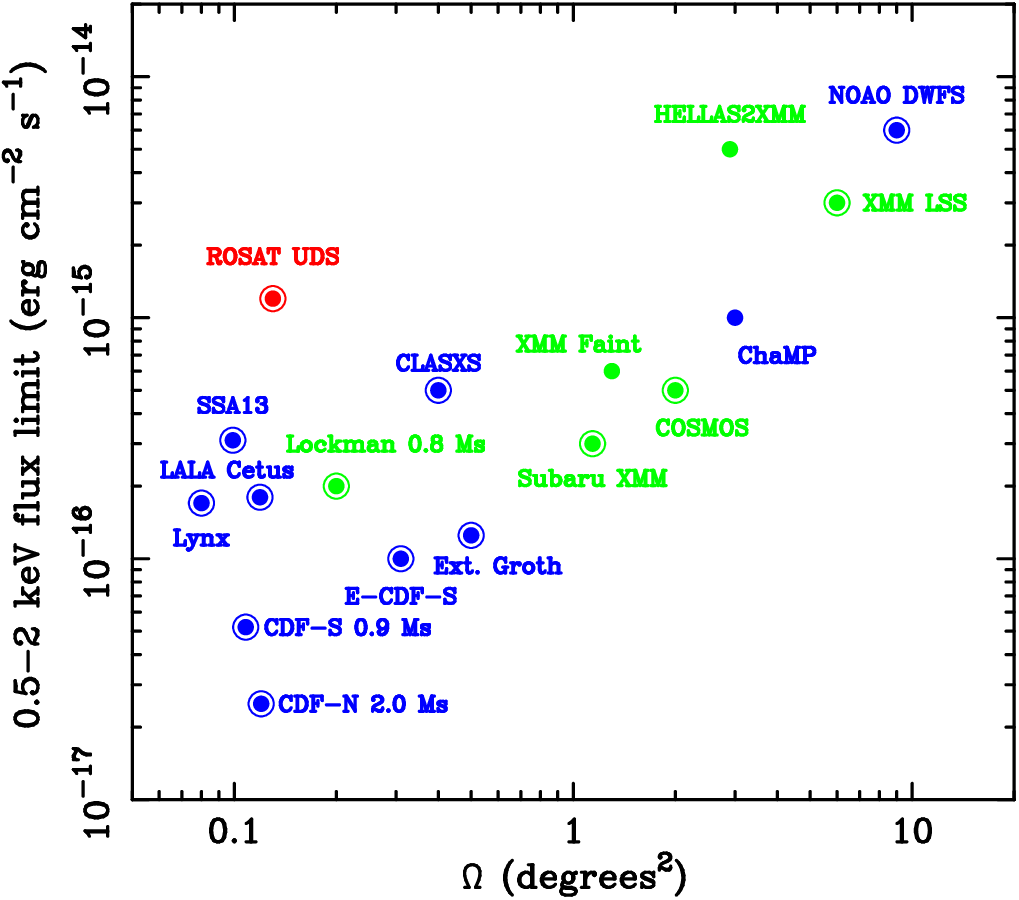}
}
\caption{
Distributions of some well-known extragalactic surveys by \chandra\
(blue), \xmm\ (green), and \rosat\ (red) in the \hbox{0.5--2~keV} flux-limit
versus solid angle, $\Omega$, plane. Circled dots denote surveys that are
contiguous. Each of the surveys shown has a range of flux limits across
its solid angle; we have generally shown the most sensitive flux limit.
This plot has been adapted from Figure~1 of Brandt \& Hasinger (2005)
to show the part of parameter space most relevant for the \hbox{E-CDF-S};
see Table~1 of Brandt \& Hasinger (2005) for references to descriptions
of many of the surveys plotted here.} \end{figure}

\clearpage

%
%%%%%%%%%%%%%%%%%%%%%%%%%%%%%%%%%%%%%%%%%%%%%%%%%%%%%%%%%%%%%%%%%%%%%%%%%%%%%%%%%%%
% Figure 2
%%%%%%%%%%%%%%%%%%%%%%%%%%%%%%%%%%%%%%%%%%%%%%%%%%%%%%%%%%%%%%%%%%%%%%%%%%%%%%%%%%
%
\begin{figure}
\figurenum{2}
\centerline{
\includegraphics[width=16.0cm]{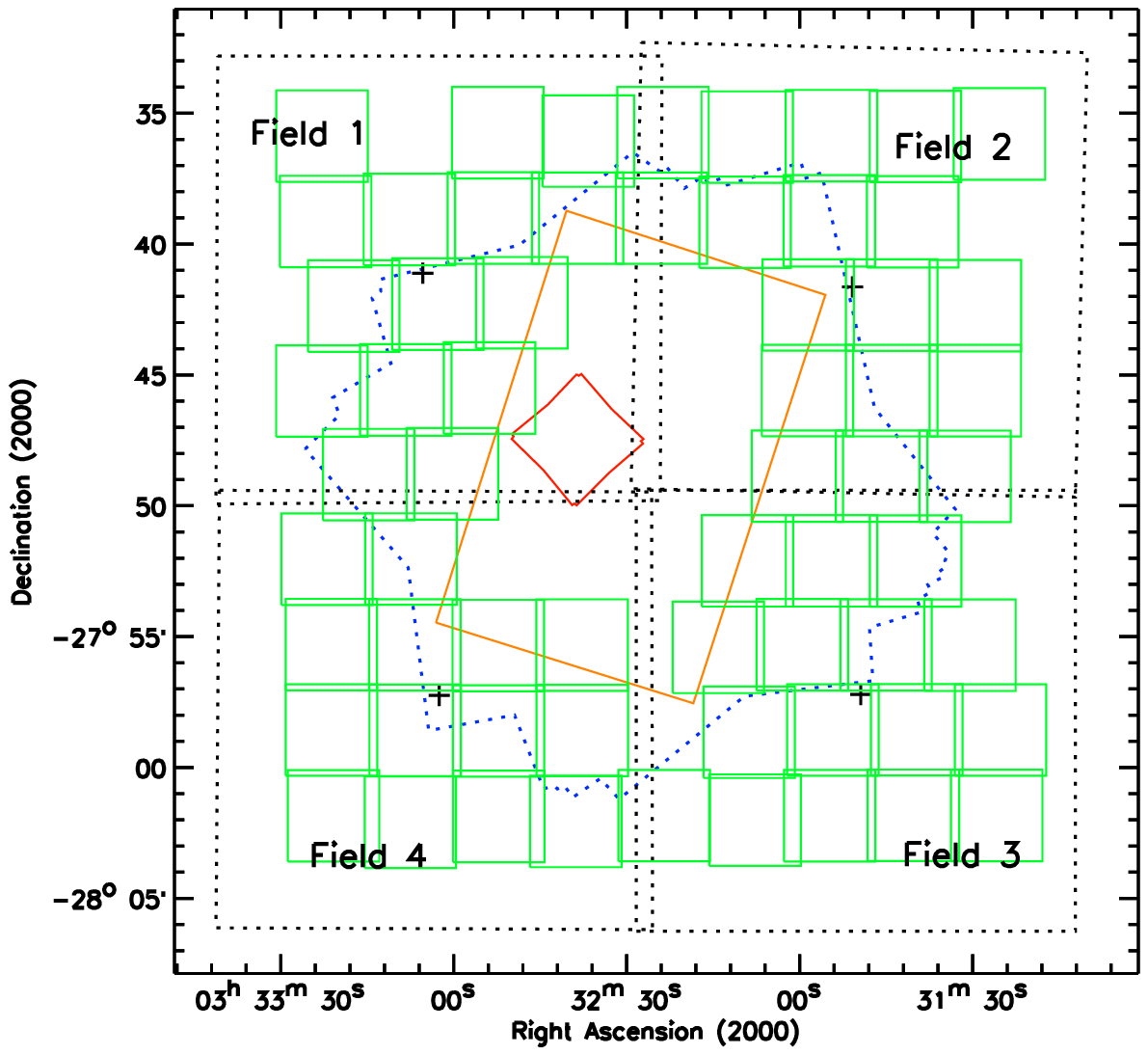}
}
\caption{
Coverage map of the \hbox{E-CDF-S} area showing the various \chandra\ (dashed
lines) and {\it HST} (solid lines) observational regions.  The \hbox{E-CDF-S}
\chandra\ observational fields are shown as four 16\farcm9 $\times$ 16\farcm9
regions (colored black in the electronic edition) which flank the CDF-S
(central polygon [colored blue in the electronic edition]).  Each observational
field is labeled in text, and the corresponding aim points are indicated as
``+'' signs within the fields.  The {\it HST} coverage includes the 63,
202$\arcsec$~$\times$~ 202$\arcsec$ square regions (colored green in the
electronic edition) from GEMS (Rix et~al.~2004), the central rectangle (colored
orange in the electronic edition) from GOODS (Giavalisco et~al.~2004), and the
central 202$\arcsec$~$\times$~ 202$\arcsec$ (colored red in the electronic
edition) region of the {\it Hubble} Ultra Deep Field (UDF; PI: S.  Beckwith).
The \spitzer\ GOODS coverage coincides with the \hst\ GOODS region (central
rectangle), and there is a substantial amount of wider field \spitzer\ coverage
either executed or approved (see $\S$~1).}
\end{figure}

\clearpage

%
%%%%%%%%%%%%%%%%%%%%%%%%%%%%%%%%%%%%%%%%%%%%%%%%%%%%%%%%%%%%%%%%%%%%%%%%%%%%%%%%%%%
% Figure 3
%%%%%%%%%%%%%%%%%%%%%%%%%%%%%%%%%%%%%%%%%%%%%%%%%%%%%%%%%%%%%%%%%%%%%%%%%%%%%%%%%%
%
\begin{figure}
\figurenum{3}
\centerline{
\includegraphics[width=16.0cm]{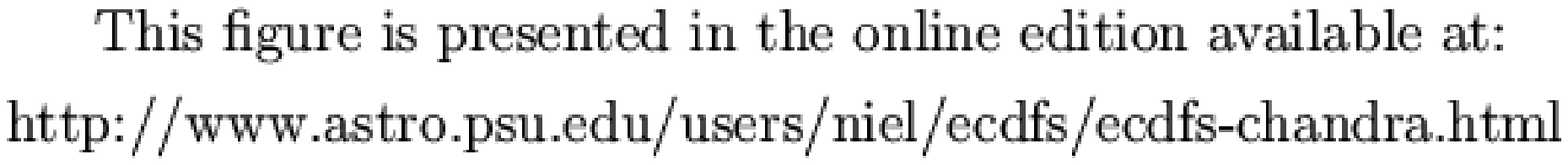}
}
\caption{
Full-band (0.5--8.0~keV) raw image of the \hbox{E-CDF-S}.  Here, all
four observational fields have been merged using the {\sc ciao} tool {\sc
merge\_all}; the {\it Chandra} aim-points for each field are marked with ``+'' symbols.
Field numbers are printed over each observational field to show their relative
locations; note the increase in background where these fields overlap.
The CDF-S, GOODS, and UDF regions are marked as per Figure~2.}
\end{figure}

\clearpage

%
%%%%%%%%%%%%%%%%%%%%%%%%%%%%%%%%%%%%%%%%%%%%%%%%%%%%%%%%%%%%%%%%%%%%%%%%%%%%%%%%%%%
% Figure 4
%%%%%%%%%%%%%%%%%%%%%%%%%%%%%%%%%%%%%%%%%%%%%%%%%%%%%%%%%%%%%%%%%%%%%%%%%%%%%%%%%%
%
\begin{figure}
\figurenum{4}
\centerline{
\includegraphics[width=16.0cm]{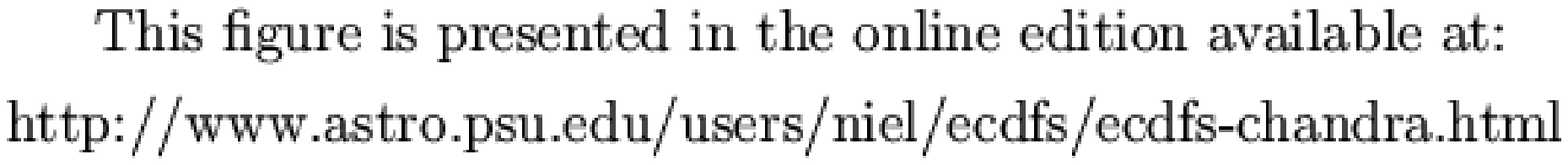}
}
\caption{Full-band (0.5--8.0~keV) adaptively smoothed and exposure-corrected image of the
\hbox{E-CDF-S}.  The image was created using the {\sc ciao} tool {\sc
csmooth}, applied to the raw-image (presented in Figure~3) at the 2.5$\sigma$
level. The grayscales are linear.  Symbols and regions have the same meaning as
those shown in Figure~3.}
\end{figure}

\clearpage

%
%%%%%%%%%%%%%%%%%%%%%%%%%%%%%%%%%%%%%%%%%%%%%%%%%%%%%%%%%%%%%%%%%%%%%%%%%%%%%%%%%%%
% Figure 5
%%%%%%%%%%%%%%%%%%%%%%%%%%%%%%%%%%%%%%%%%%%%%%%%%%%%%%%%%%%%%%%%%%%%%%%%%%%%%%%%%%
%
\begin{figure}
\figurenum{5}
\centerline{
\includegraphics[width=16.0cm]{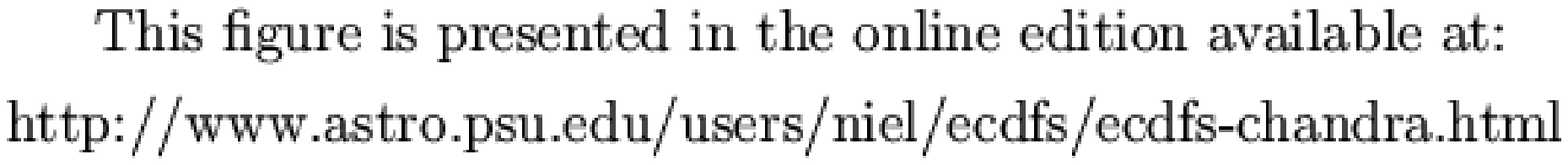}
}
\caption{
Full-band exposure map of the \hbox{E-CDF-S}.  The
grayscales are linear with the darkest areas corresponding to the highest
effective exposure times (the high effective exposure times between fields is
due to overlap of observations).  Note the chip gaps in white running between
the four ACIS-I CCDs.  Symbols and regions
have the same meaning as those shown in Figure~3.}
\end{figure}

%
%%%%%%%%%%%%%%%%%%%%%%%%%%%%%%%%%%%%%%%%%%%%%%%%%%%%%%%%%%%%%%%%%%%%%%%%%%%%%%%%%%%
% Figure 6
%%%%%%%%%%%%%%%%%%%%%%%%%%%%%%%%%%%%%%%%%%%%%%%%%%%%%%%%%%%%%%%%%%%%%%%%%%%%%%%%%%
%

\begin{figure}
\figurenum{6}
\centerline{
\includegraphics[width=8.0cm]{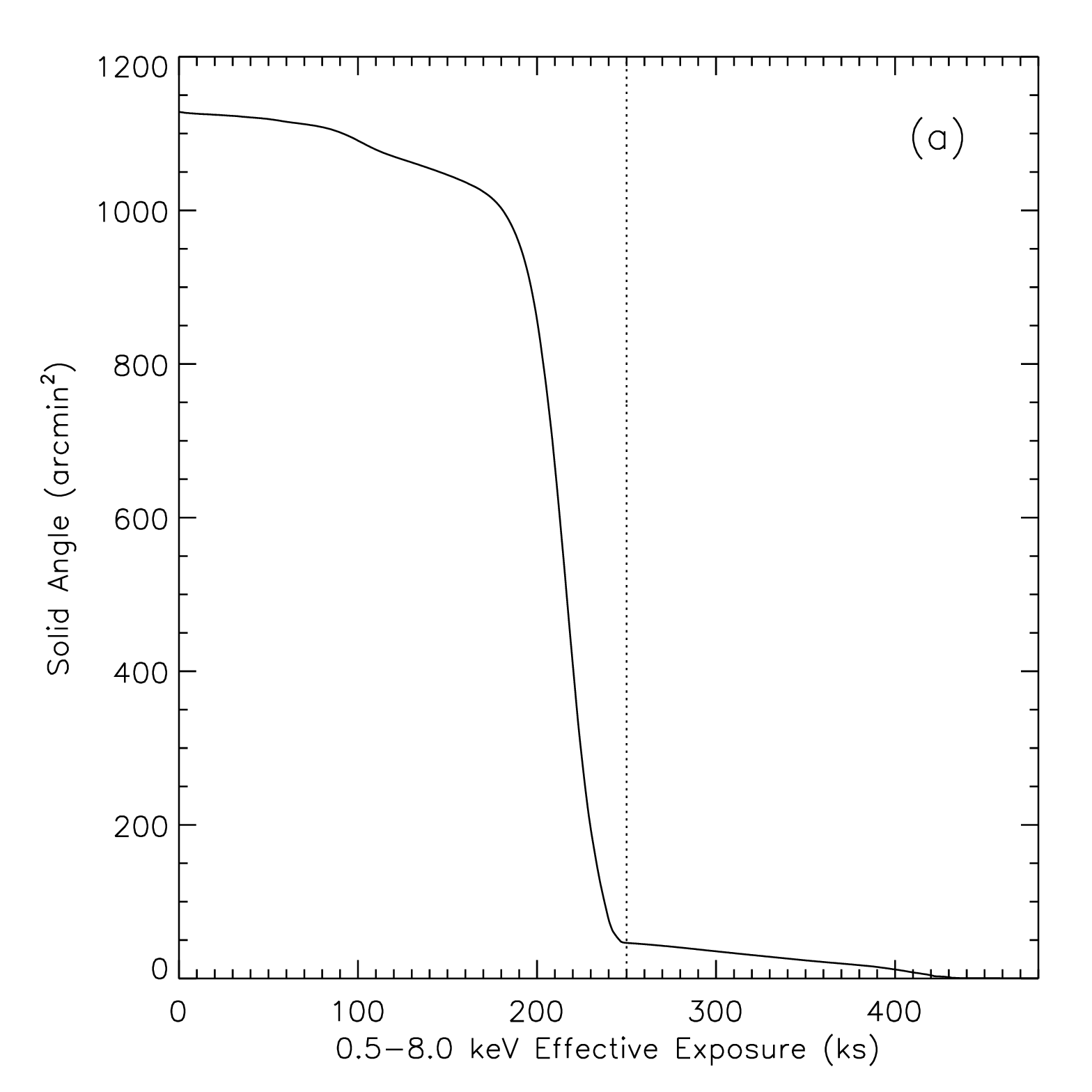}
\hfill
\includegraphics[width=8.0cm]{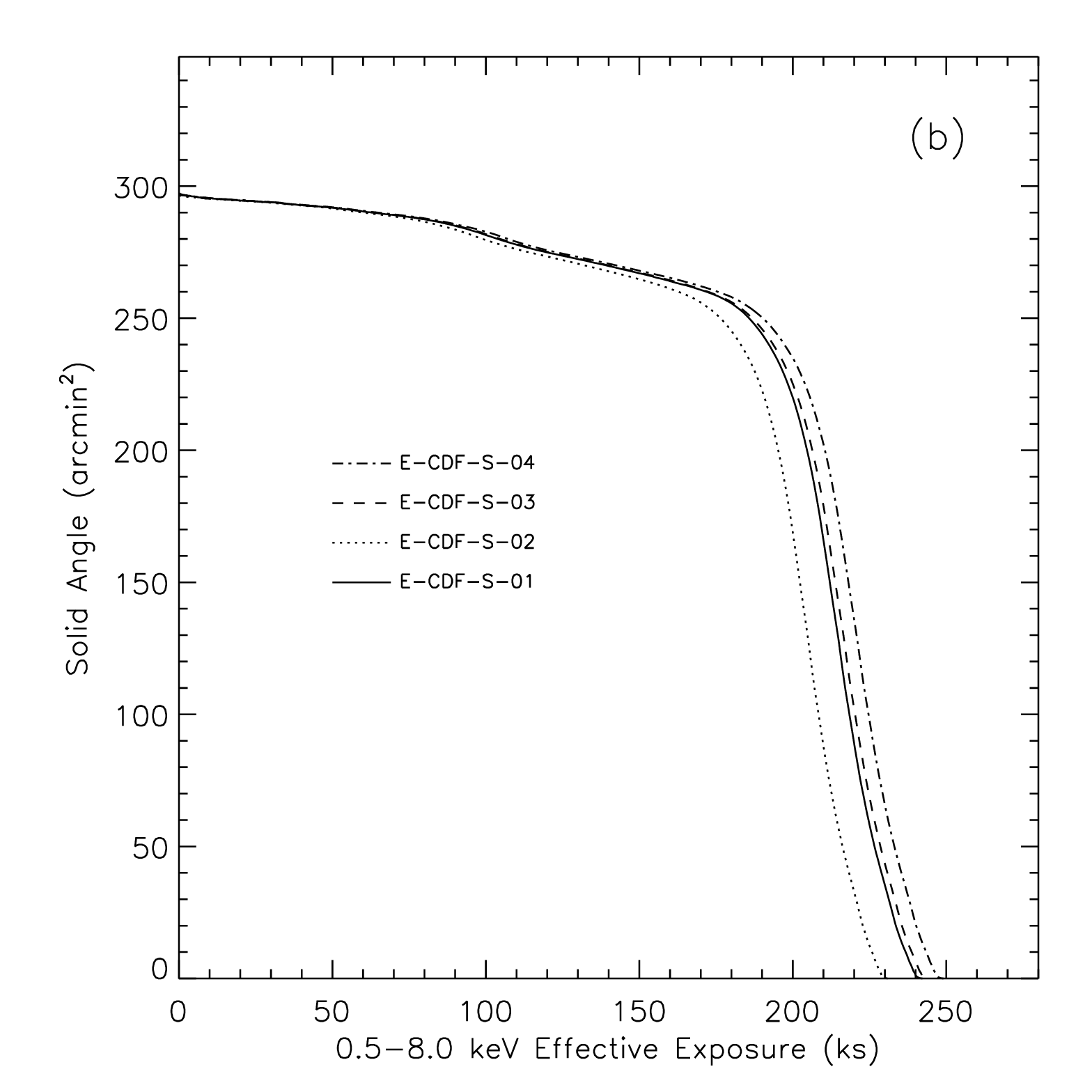}
}
\caption{
{\bf (a)} Amount of survey solid angle having at least a given amount of
effective exposure in the full-band exposure map for the entire \hbox{E-CDF-S}.
The ``tail'' with exposure times $>$~250~ks (right of the vertical line)
corresponds to regions where observational fields overlap
(see Figure~5).  {\bf (b)} Amount of solid angle having at least a given amount
of effective full-band exposure for each of the four fields.  
}
\end{figure}

%
%%%%%%%%%%%%%%%%%%%%%%%%%%%%%%%%%%%%%%%%%%%%%%%%%%%%%%%%%%%%%%%%%%%%%%%%%%%%%%%%%%%
% Figure 7
%%%%%%%%%%%%%%%%%%%%%%%%%%%%%%%%%%%%%%%%%%%%%%%%%%%%%%%%%%%%%%%%%%%%%%%%%%%%%%%%%%
%
\begin{figure}
\figurenum{7}
\centerline{
\includegraphics[width=16.0cm]{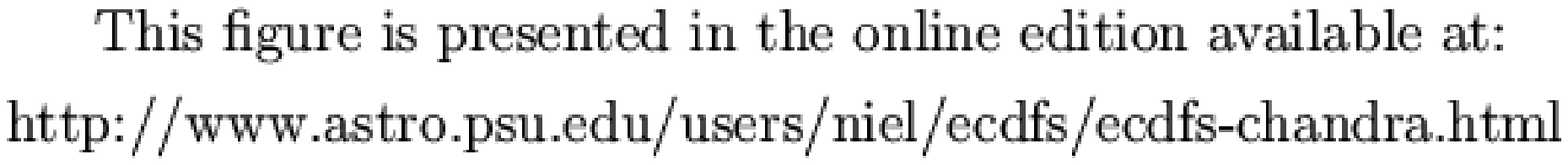}
}
\caption{
\chandra\ ``false-color'' image of the \hbox{E-CDF-S}.
This image has been constructed from the \hbox{0.5--2.0~keV} (red), \hbox{2--4~keV}
(green), and \hbox{4--8~keV} (blue) exposure-corrected adaptively smoothed
images discussed in $\S$~3.1.  } 
\end{figure}

%
%%%%%%%%%%%%%%%%%%%%%%%%%%%%%%%%%%%%%%%%%%%%%%%%%%%%%%%%%%%%%%%%%%%%%%%%%%%%%%%%%%%
% Figure 8
%%%%%%%%%%%%%%%%%%%%%%%%%%%%%%%%%%%%%%%%%%%%%%%%%%%%%%%%%%%%%%%%%%%%%%%%%%%%%%%%%%
%
\begin{figure}
\figurenum{8}
\centerline{
\includegraphics[width=10.0cm]{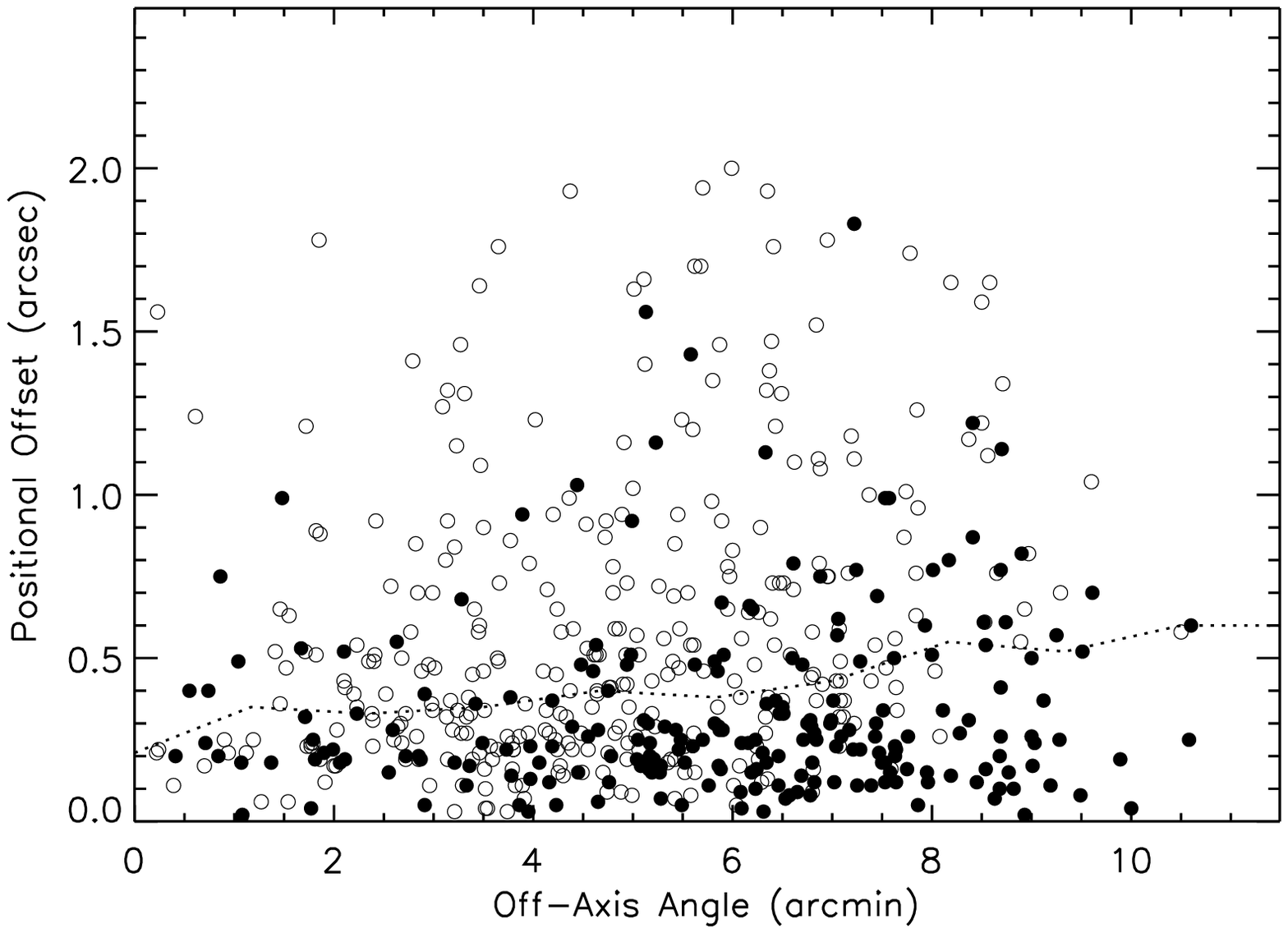}
}
\caption{
Positional offset vs. off-axis angle (computed in each observational field) for
sources in the main {\it Chandra} catalog that were matched to optical sources
from the WFI $R$-band image to within 2$\farcs$5.  Open circles are {\it
Chandra} sources with $<$ 50 full-band
counts, and filled circles are {\it Chandra} sources with $\ge$~50 full-band
counts.  The dotted curve shows the running median for all sources. These data were
used to determine the positional uncertainties of the {\it Chandra} sources;
see $\S$~3.3.1.
} 
\end{figure}

%
%%%%%%%%%%%%%%%%%%%%%%%%%%%%%%%%%%%%%%%%%%%%%%%%%%%%%%%%%%%%%%%%%%%%%%%%%%%%%%%%%%%
% Figure 9
%%%%%%%%%%%%%%%%%%%%%%%%%%%%%%%%%%%%%%%%%%%%%%%%%%%%%%%%%%%%%%%%%%%%%%%%%%%%%%%%%%
%
\begin{figure}
\figurenum{9}
\centerline{
\includegraphics[width=12.0cm]{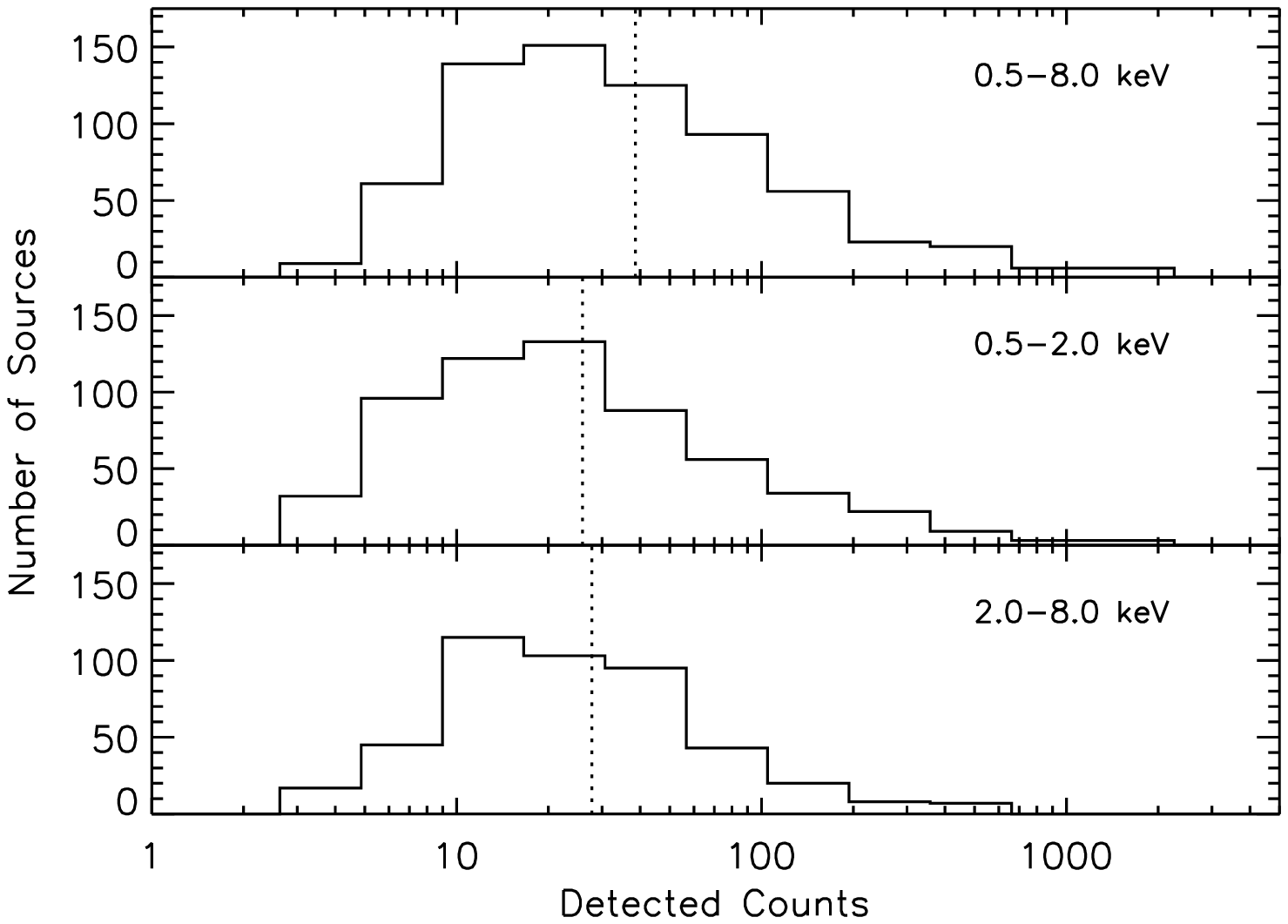}
}
\caption{
Histograms showing the distributions of detected source counts for sources in the main
{\it Chandra} catalog in the full ({\it top}), soft ({\it middle}), and hard ({\it
bottom}) bands.  Sources with upper limits have not been included in this plot.
The vertical dotted lines indicate median numbers of counts (see Table~4).  
}
\end{figure}

%
%%%%%%%%%%%%%%%%%%%%%%%%%%%%%%%%%%%%%%%%%%%%%%%%%%%%%%%%%%%%%%%%%%%%%%%%%%%%%%%%%%%
% Figure 10
%%%%%%%%%%%%%%%%%%%%%%%%%%%%%%%%%%%%%%%%%%%%%%%%%%%%%%%%%%%%%%%%%%%%%%%%%%%%%%%%%%
%
\begin{figure}
\figurenum{10}
\centerline{
\includegraphics[width=12.0cm]{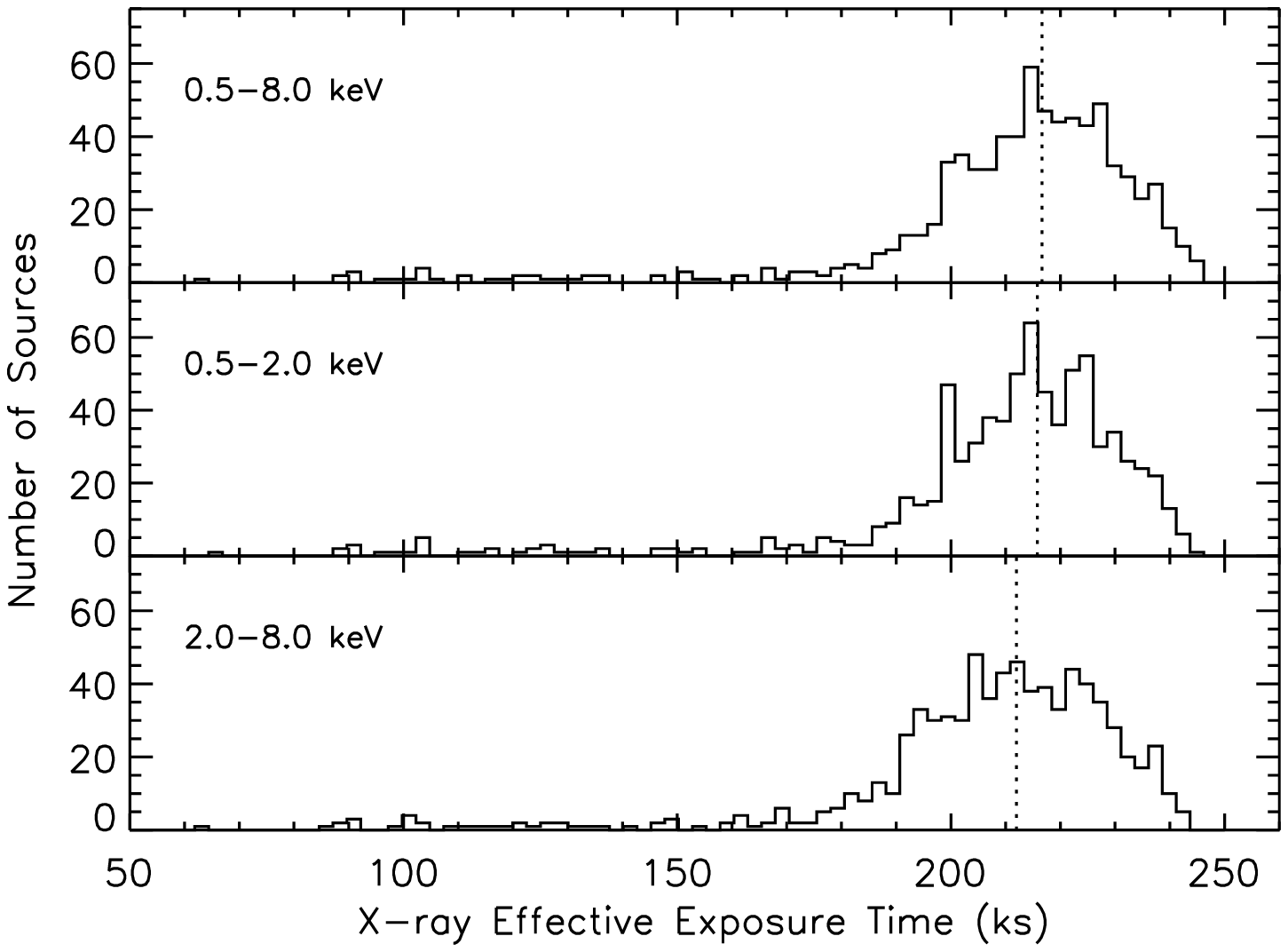}
}
\caption{
Histograms showing the distributions of effective exposure time for the sources
in the main {\it Chandra} catalog in the full ({\it top}), soft ({\it middle}),
and hard ({\it bottom}) bands.  The vertical dotted lines indicate the median
effective exposure times of \fbexp, \sbexp, \hbexp~ks for the full, soft, and hard bands,
respectively.}
\end{figure}

%
%%%%%%%%%%%%%%%%%%%%%%%%%%%%%%%%%%%%%%%%%%%%%%%%%%%%%%%%%%%%%%%%%%%%%%%%%%%%%%%%%%%
% Figure 11
%%%%%%%%%%%%%%%%%%%%%%%%%%%%%%%%%%%%%%%%%%%%%%%%%%%%%%%%%%%%%%%%%%%%%%%%%%%%%%%%%%
%
\begin{figure}
\figurenum{11}
\centerline{
\includegraphics[width=12.0cm]{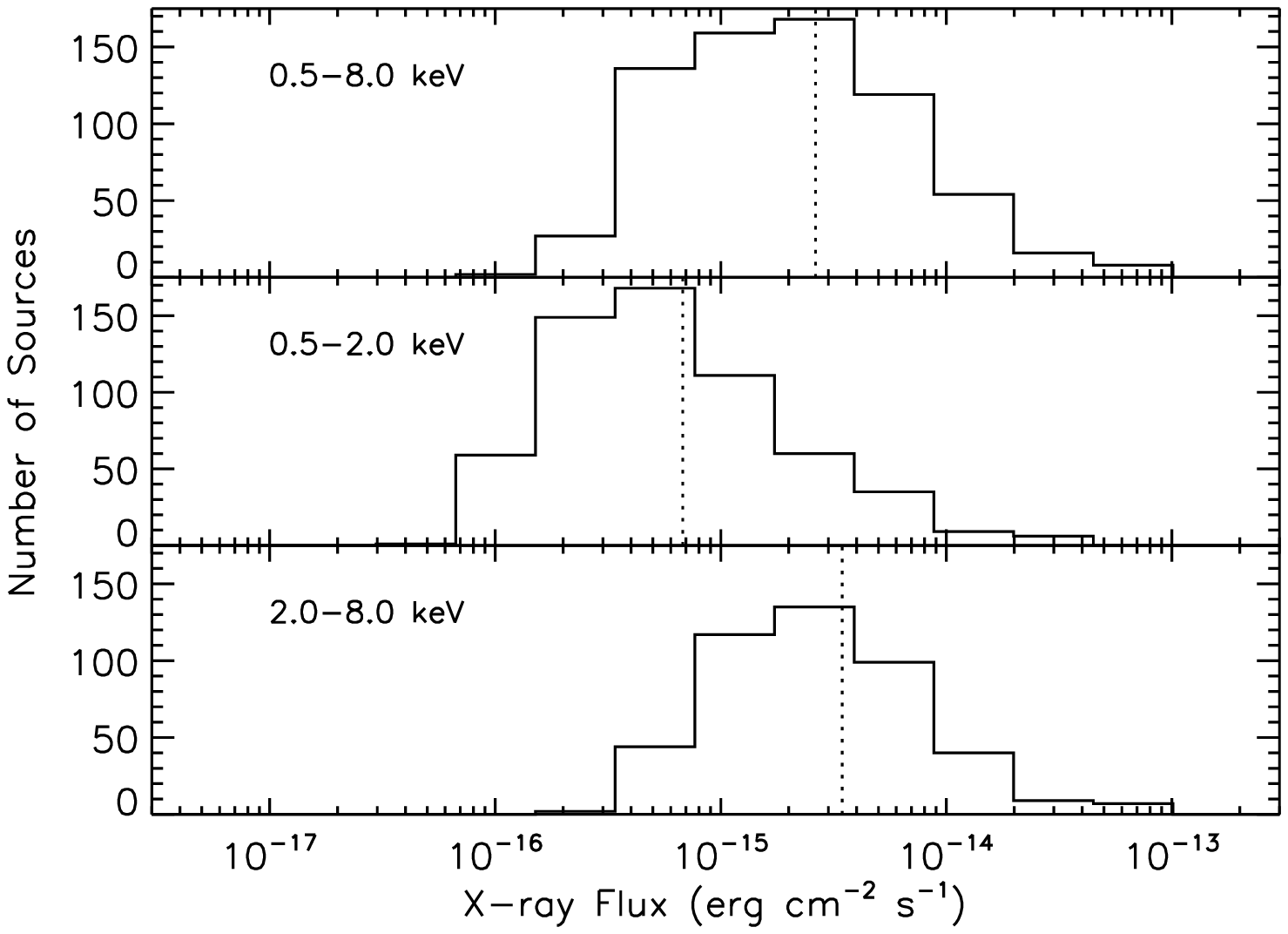}
}
\caption{
Histograms showing the distributions of \hbox{X-ray} fluxes for sources in the main
{\it Chandra} catalog in the full ({\it top}), soft ({\it middle}), and hard ({\it
bottom}) bands.  Sources with upper limits have not been included in this plot.
The vertical dotted lines indicate the median fluxes of \fbmean, \sbmean, and
\hbox{\hbmean\ $\times$~10$^{-16}$ \flux} for the full, soft, and hard bands, respectively.} 
\end{figure}

%
%%%%%%%%%%%%%%%%%%%%%%%%%%%%%%%%%%%%%%%%%%%%%%%%%%%%%%%%%%%%%%%%%%%%%%%%%%%%%%%%%%%
% Figure 12
%%%%%%%%%%%%%%%%%%%%%%%%%%%%%%%%%%%%%%%%%%%%%%%%%%%%%%%%%%%%%%%%%%%%%%%%%%%%%%%%%%
%
\begin{figure}
\figurenum{12}
\centerline{
\includegraphics[width=15.0cm]{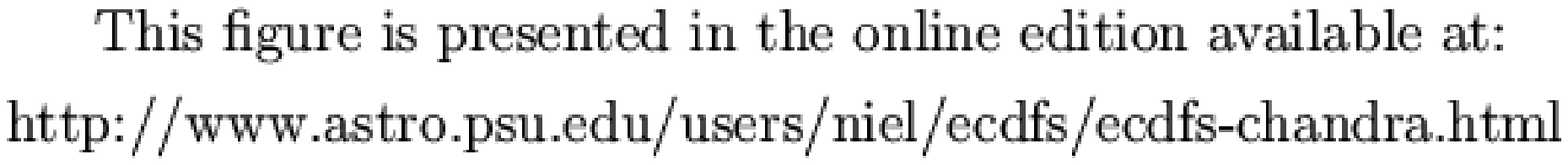}
}
\caption{
WFI $R$-band postage-stamp images for the sources in the main {\it Chandra}
catalog with full-band adaptively smoothed \hbox{X-ray} contours overlaid.  The
contours are logarithmic in scale and range from \hbox{$\approx$0.03--30\%} of the
maximum pixel value.  Note that for sources with few full-band counts, {\sc
csmooth} has suppressed the observable emission in the adaptively smoothed
images and therefore no \hbox{X-ray} contours are observed for these sources.
The label at the top of each image gives the source name, which is composed of
the source coordinates, while numbers at the bottom left and right hand corners
correspond to the source number (see column~1 of Table~1) and full-band source
counts, respectively.  Each image is $\approx$24\farcs6 on a side, and the
source of interest is always located at the center of the image. Only one of
the thirteen pages of cutouts is included here; all thirteen pages are
available at the \hbox{E-CDF-S} website
(http://www.astro.psu.edu/users/niel/ecdfs/ecdfs-chandra.html).} \end{figure}

%
%%%%%%%%%%%%%%%%%%%%%%%%%%%%%%%%%%%%%%%%%%%%%%%%%%%%%%%%%%%%%%%%%%%%%%%%%%%%%%%%%%%
% Figure 13
%%%%%%%%%%%%%%%%%%%%%%%%%%%%%%%%%%%%%%%%%%%%%%%%%%%%%%%%%%%%%%%%%%%%%%%%%%%%%%%%%%
%
\begin{figure}
\figurenum{13}
\centerline{
\includegraphics[width=16.0cm]{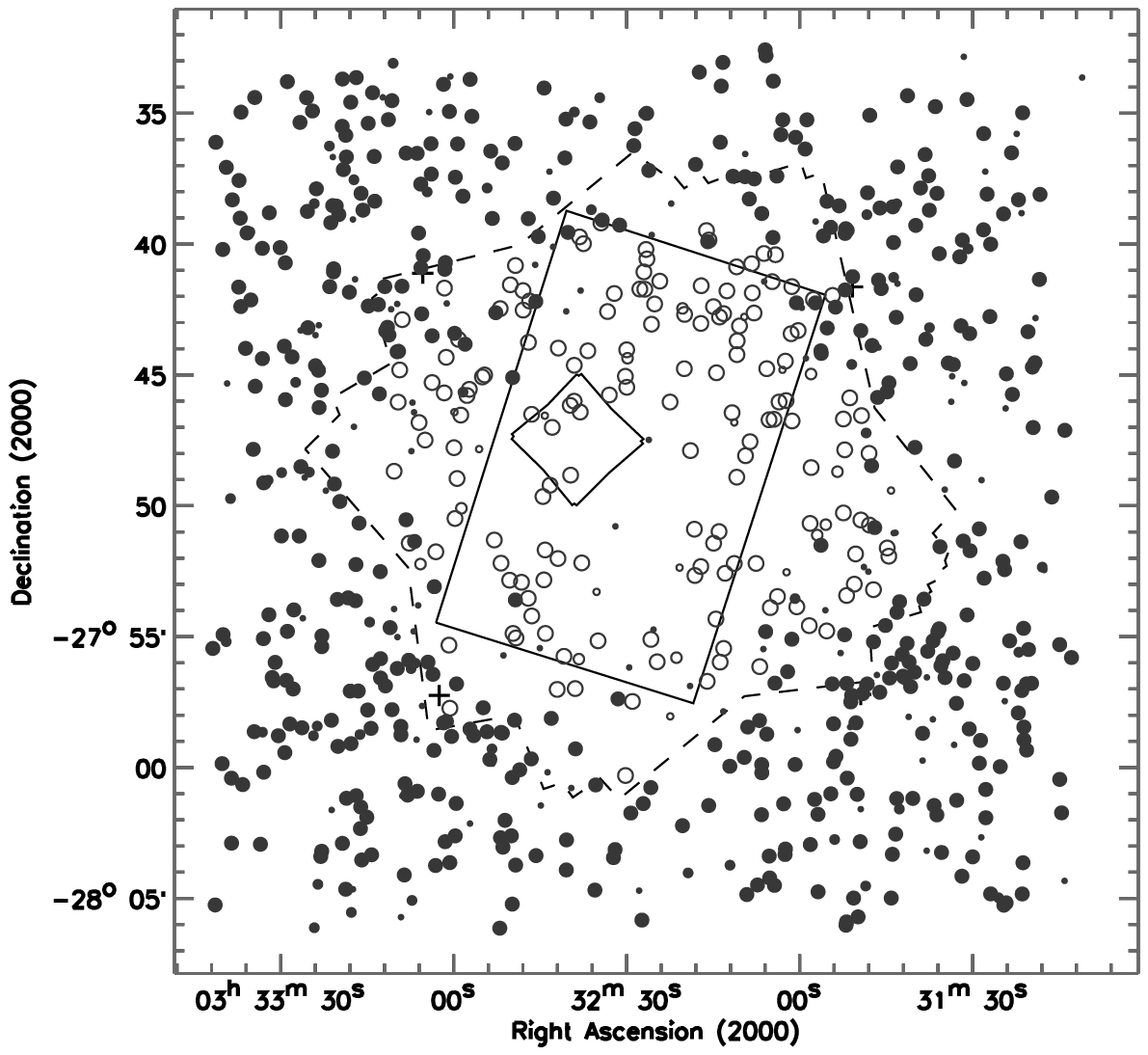}
}
\caption{
Positions of the sources in the main {\it Chandra} catalog.  The
regions have the same meaning as those given in Figure~3.  New
\hbox{X-ray} sources are shown here as filled circles while sources that were
previously detected in the $\approx$1~Ms CDF-S are shown as open circles.
Large, medium, and small circles correspond to sources with {\sc wavdetect}
false-positive probability $\le$~1~$\times$~10$^{-8}$,
$\le$~1~$\times$~10$^{-7}$, and $\le$~1~$\times$~10$^{-6}$, respectively.}
\end{figure}

%
%%%%%%%%%%%%%%%%%%%%%%%%%%%%%%%%%%%%%%%%%%%%%%%%%%%%%%%%%%%%%%%%%%%%%%%%%%%%%%%%%%%
% Figure 14
%%%%%%%%%%%%%%%%%%%%%%%%%%%%%%%%%%%%%%%%%%%%%%%%%%%%%%%%%%%%%%%%%%%%%%%%%%%%%%%%%%
%
\begin{figure}
\figurenum{14}
\centerline{
\includegraphics[width=12.0cm]{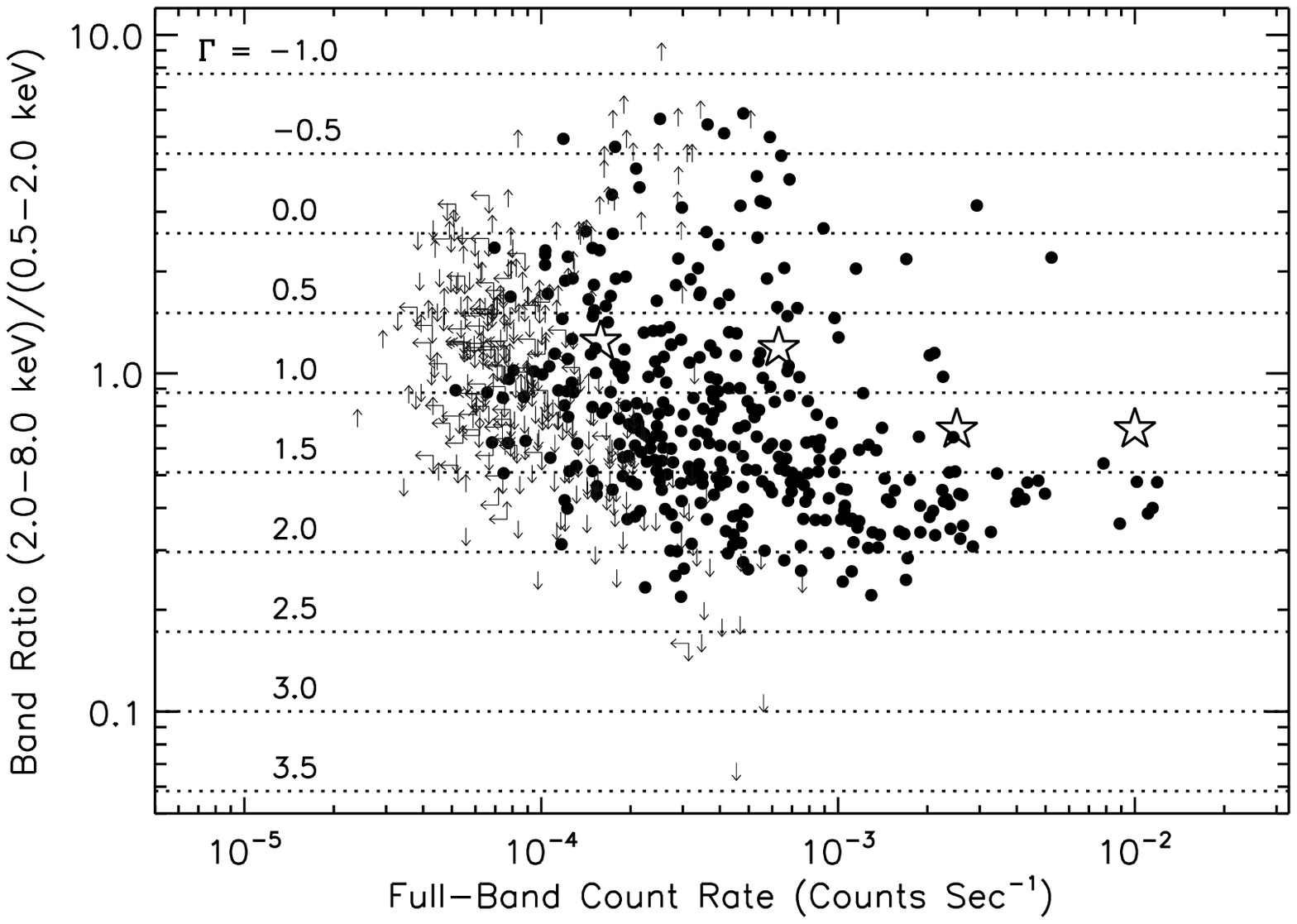}
}
\caption{
Band ratio as a function of full-band count rate for the sources in the main
{\it Chandra} catalog.  Small solid dots show sources detected in both the soft
and hard bands. Plain arrows show sources detected in only one of these two
bands with the arrows indicating upper and lower limits; sources detected in
only the full band cannot be plotted. The open stars show average band ratios
as a function of full-band count rate.
Horizontal dotted lines are labeled with the photon indices that correspond to
a given band ratio assuming only Galactic absorption (these were determined
using the CXC's Portable, Interactive, Multi-Mission Simulator; PIMMS).}
\end{figure}

%
%%%%%%%%%%%%%%%%%%%%%%%%%%%%%%%%%%%%%%%%%%%%%%%%%%%%%%%%%%%%%%%%%%%%%%%%%%%%%%%%%%%
% Figure 15
%%%%%%%%%%%%%%%%%%%%%%%%%%%%%%%%%%%%%%%%%%%%%%%%%%%%%%%%%%%%%%%%%%%%%%%%%%%%%%%%%%
%
\begin{figure}
\figurenum{15}
\centerline{
\includegraphics[width=12.0cm]{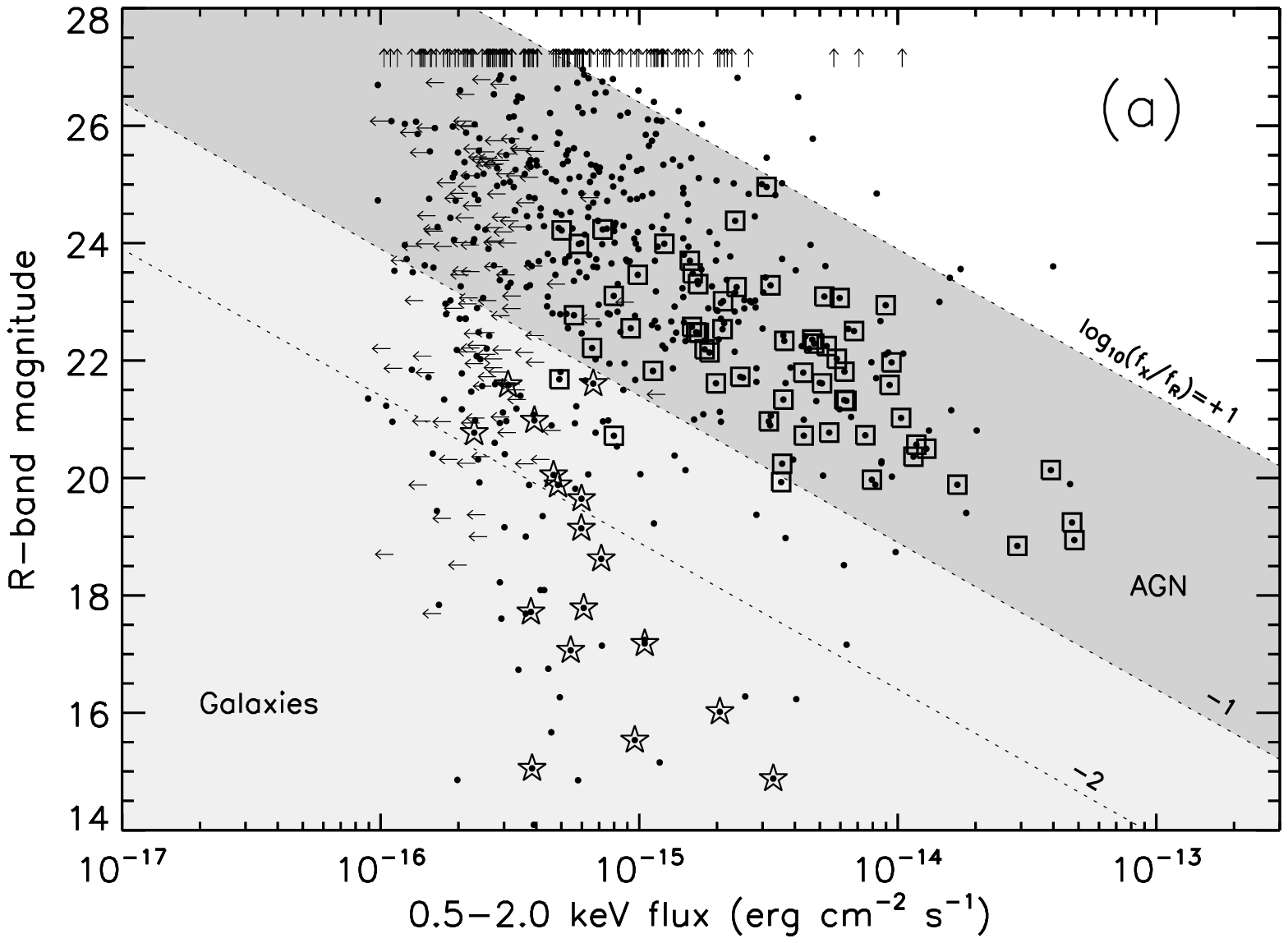}
}
\centerline{
\includegraphics[width=12.0cm]{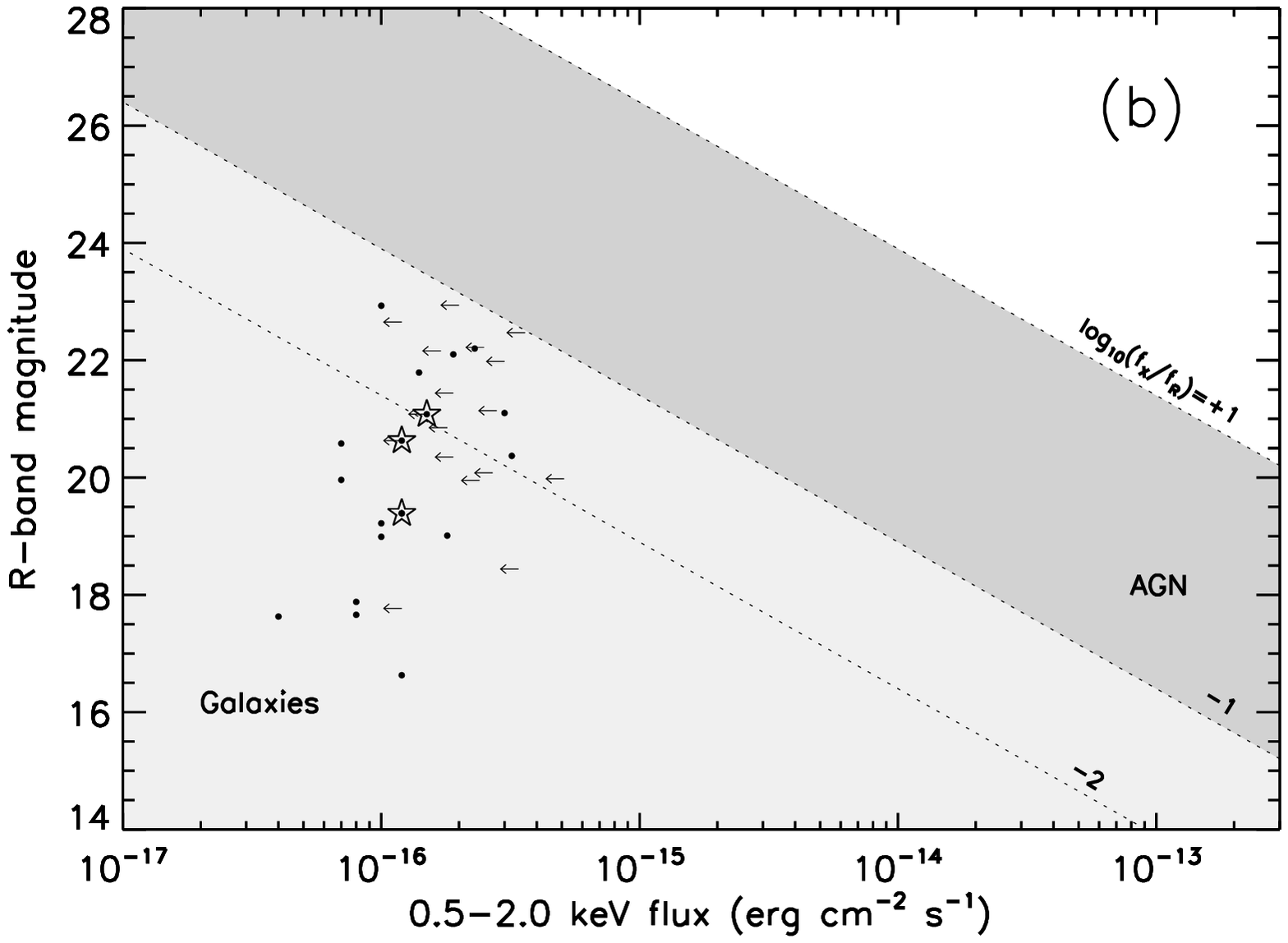}
}
\caption{
{\bf (a)} WFI $R$-band magnitude versus soft-band flux for sources in the main {\it
Chandra} catalog.  Open-star symbols indicate Galactic stars identified using
the optical spectrophotometric COMBO-17 survey (Wolf et al.~2004).  Diagonal
lines indicate constant flux ratios.  Sources that were not detected in the
soft band that were detected in at least one of the full and hard bands are
plotted here as upper limits.  The shaded regions show the approximate flux
ratios for AGNs and galaxies (dark and light, respectively); the sixty-one AGNs
with reliable COMBO-17 identifications are plotted as squares. {\bf (b)}  WFI $R$-band
magnitude versus soft-band flux for sources in the optically bright
supplementary catalog.  Note that many of these sources have
\hbox{X-ray}-to-optical flux ratios expected for normal and starburst
galaxies.} \end{figure}

%
%%%%%%%%%%%%%%%%%%%%%%%%%%%%%%%%%%%%%%%%%%%%%%%%%%%%%%%%%%%%%%%%%%%%%%%%%%%%%%%%%%%
% Figure 16
%%%%%%%%%%%%%%%%%%%%%%%%%%%%%%%%%%%%%%%%%%%%%%%%%%%%%%%%%%%%%%%%%%%%%%%%%%%%%%%%%%
%
\begin{figure}
\figurenum{16}
\centerline{
\includegraphics[width=16.0cm]{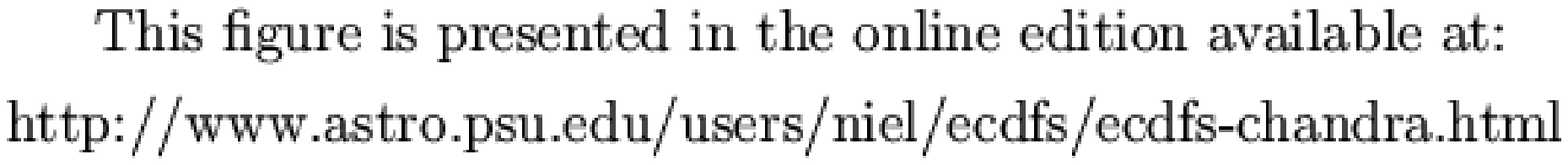}
}
\caption{Full-band background map of the \hbox{E-CDF-S}.  This background map
has been created following $\S$~4.  Symbols and regions have the same meaning as those
given in Figure~3.  }
\end{figure}

%
%%%%%%%%%%%%%%%%%%%%%%%%%%%%%%%%%%%%%%%%%%%%%%%%%%%%%%%%%%%%%%%%%%%%%%%%%%%%%%%%%%%
% Figure 17
%%%%%%%%%%%%%%%%%%%%%%%%%%%%%%%%%%%%%%%%%%%%%%%%%%%%%%%%%%%%%%%%%%%%%%%%%%%%%%%%%%
%
\begin{figure}
\figurenum{17}
\centerline{
\includegraphics[width=16.0cm]{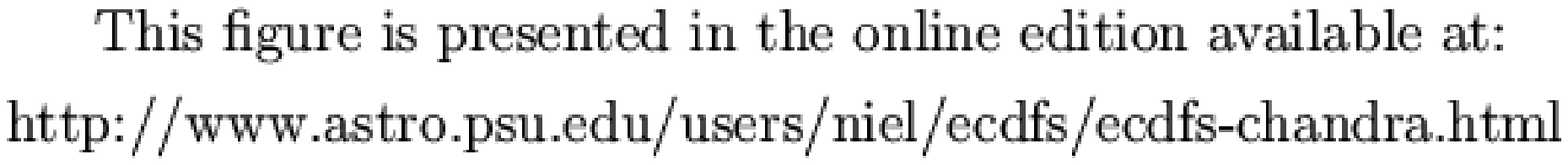}
}
\caption{Full-band sensitivity map of the \hbox{E-CDF-S}.  This
sensitivity map has been created following $\S$~4.  Symbols and regions have
the same meaning as those given in Figure~3.  Black, dark gray, light gray, and
white areas correspond to flux-limits of $<$3~$\times$~10$^{-16}$
\flux, \hbox{3--7.8~$\times$~10$^{-16}$} \flux, \hbox{7.8--20~$\times$~10$^{-16}$}
\flux, and $>$ 20~$\times$~10$^{-16}$ \flux, respectively. The central dashed
circle ($\approx$6\arcmin\ radius) shows the approximate region of the
$\approx$1~Ms CDF-S where the full band flux-limit is $<$3~$\times$~10$^{-16}$
\flux.  Note the most sensitive regions of the E-CDF-S exposure lie just
outside the $\approx$1~Ms CDF-S exposure.}
\end{figure}

%
%%%%%%%%%%%%%%%%%%%%%%%%%%%%%%%%%%%%%%%%%%%%%%%%%%%%%%%%%%%%%%%%%%%%%%%%%%%%%%%%%%%
% Figure 18
%%%%%%%%%%%%%%%%%%%%%%%%%%%%%%%%%%%%%%%%%%%%%%%%%%%%%%%%%%%%%%%%%%%%%%%%%%%%%%%%%%
%
\begin{figure}
\figurenum{18}
\centerline{
\includegraphics[width=16.0cm]{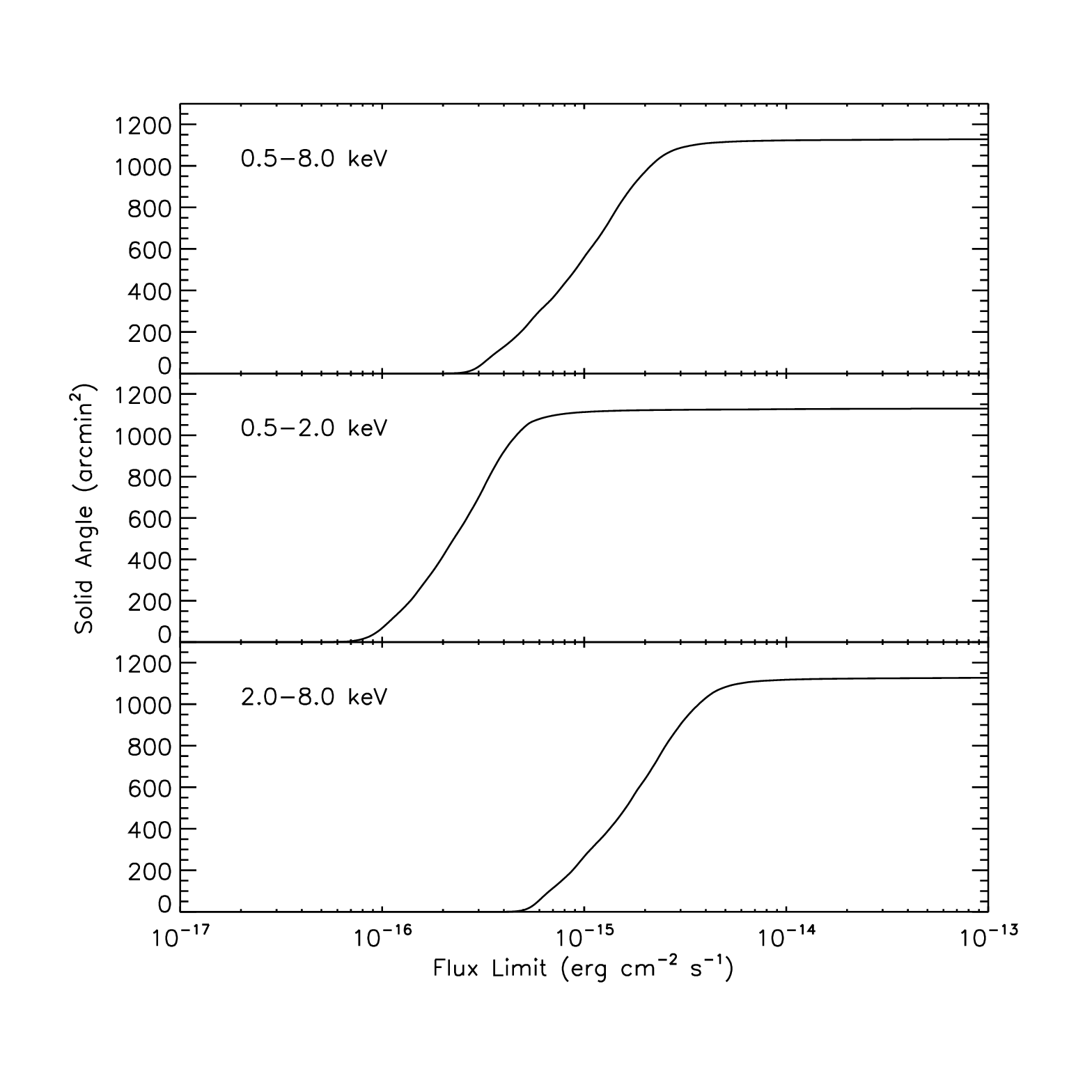}
}
\caption{Solid angle vs. flux limit for the full ({\it top}), soft
({\it middle}), and hard ({\it bottom}) bands, determined following $\S$~4.
The average flux limits (averaged over the four observational fields) at the
aim points are \fblimit~\flux\ (full-band), \sblimit~\flux\ (soft-band), and
\hblimit~\flux\ (hard-band).  } \end{figure}

%
%%%%%%%%%%%%%%%%%%%%%%%%%%%%%%%%%%%%%%%%%%%%%%%%%%%%%%%%%%%%%%%%%%%%%%%%%%%%%%%%%%%
% Figure 19
%%%%%%%%%%%%%%%%%%%%%%%%%%%%%%%%%%%%%%%%%%%%%%%%%%%%%%%%%%%%%%%%%%%%%%%%%%%%%%%%%%
%

\begin{figure}
\figurenum{19}
\centerline{
\includegraphics[width=12.0cm]{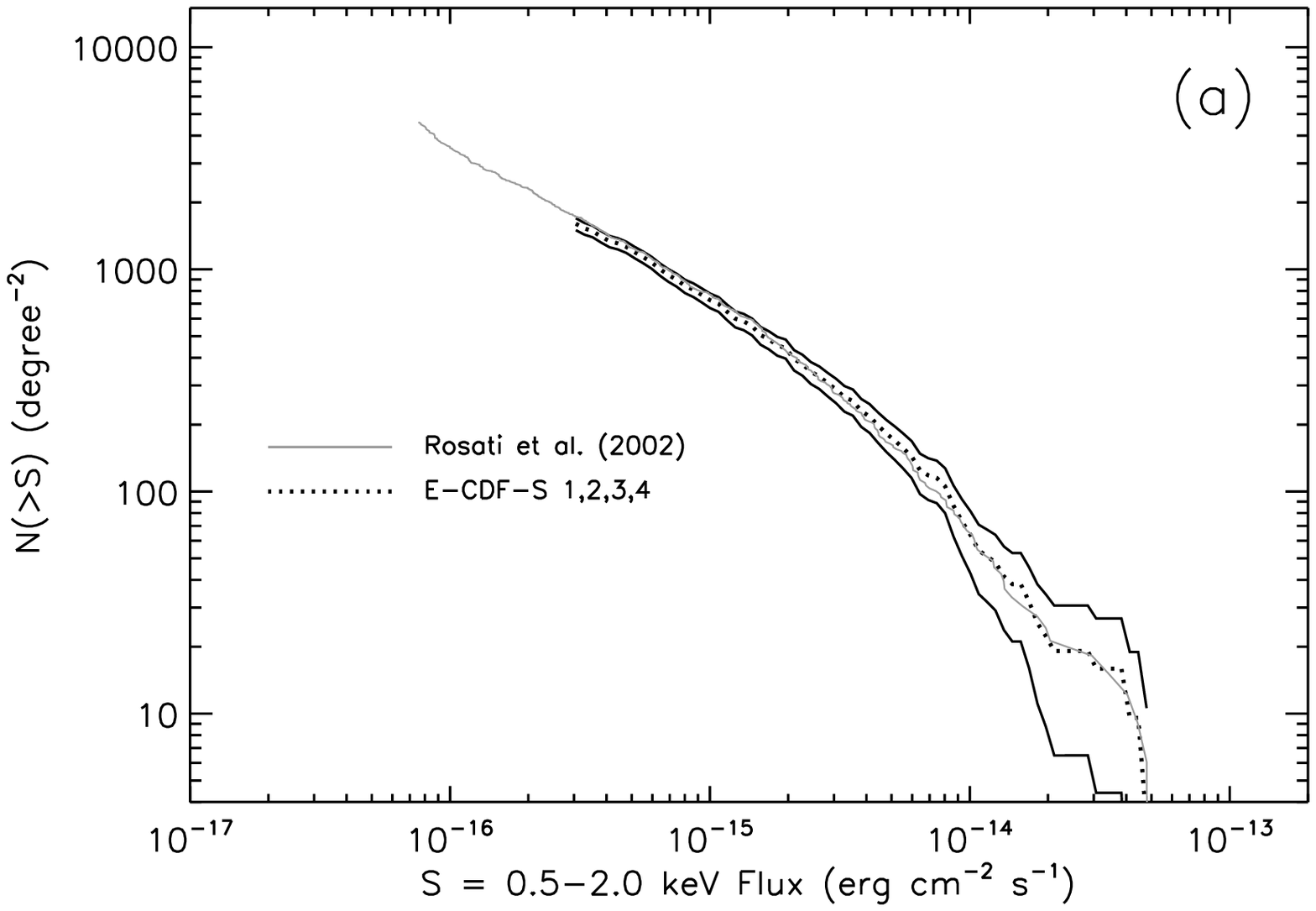}
}
\centerline{
\includegraphics[width=12.0cm]{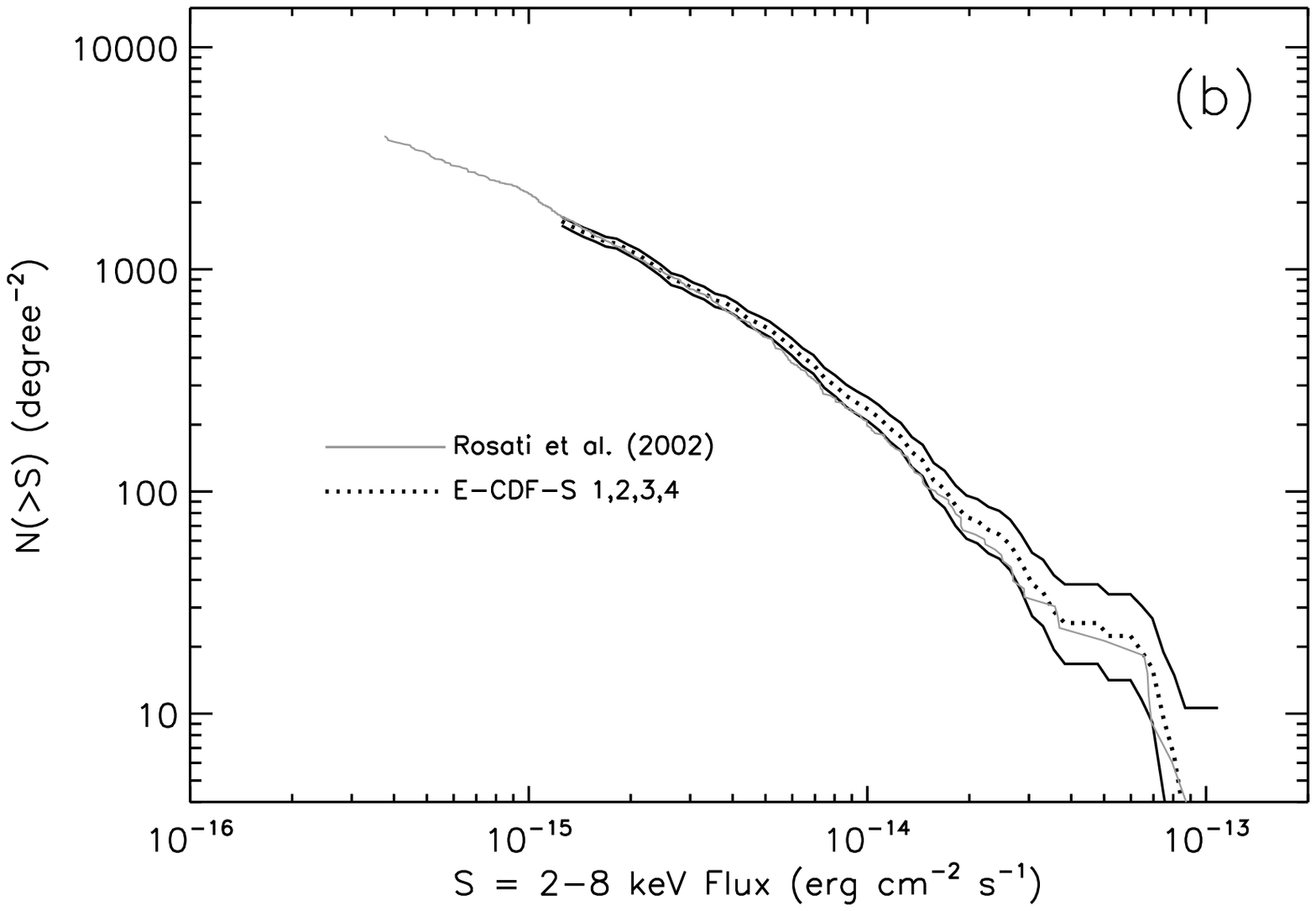}
}
\caption{
Number of sources, $N(>S)$, brighter than a given flux, $S$, for the {\bf (a)}
soft band and {\bf (b)} hard band.  In both figures, data from our main
\chandra\ catalog are plotted as black dotted curves with the 1$\sigma$ errors
(computed following Gehrels et~al.~1986) plotted as black solid curves.  In both
figures, the observed number counts for the $\approx$1~Ms \hbox{CDF-S} (adapted
from Rosati et~al. 2002) are plotted in gray for comparison.   }
\end{figure}

%
%%%%%%%%%%%%%%%%%%%%%%%%%%%%%%%%%%%%%%%%%%%%%%%%%%%%%%%%%%%%%%%%%%%%%%%%%%%%%%%%%%%
% Figure 20
%%%%%%%%%%%%%%%%%%%%%%%%%%%%%%%%%%%%%%%%%%%%%%%%%%%%%%%%%%%%%%%%%%%%%%%%%%%%%%%%%%
%

\begin{figure}
\figurenum{20}
\centerline{
\includegraphics[width=8cm]{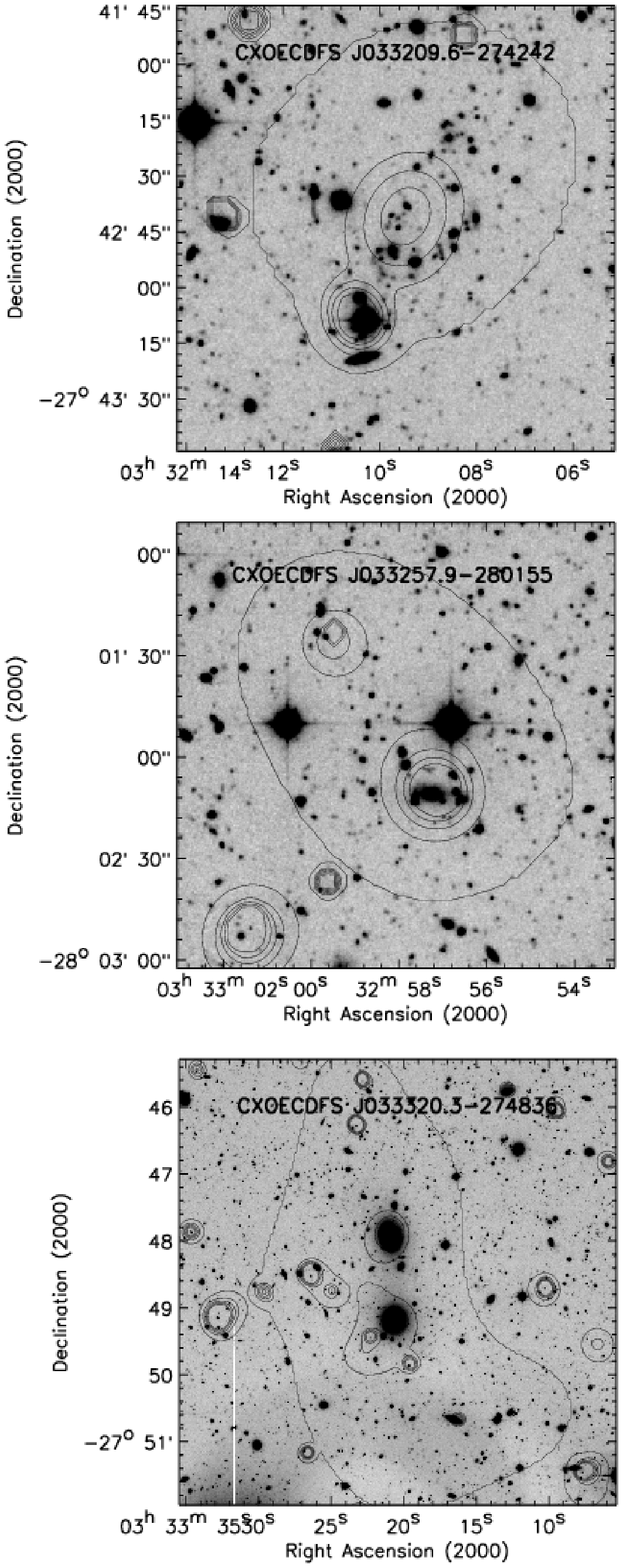}
}
\caption{WFI $R$-band images with adaptively smoothed 0.5--2.0~keV \hbox{X-ray}
contours of the spatially extended X-ray sources \hbox{CXOECDFS
J033209.6$-$274242}, \hbox{CXOECDFS J033257.9$-$280155}, and \hbox{CXOECDFS
J033320.3$-$274836}.  Contours are at 10\%, 30\%, 50\%, 70\%, and 90\% of the
maximum pixel value.  Note that each panel has a different size.} 
\end{figure}

\end{document}